\documentclass{aa}  
\usepackage{graphicx}
\usepackage{txfonts}
\usepackage{hyperref}
\hypersetup{colorlinks=true,linkcolor=blue,urlcolor=blue,citecolor=blue}
\usepackage[usenames,dvipsnames]{xcolor}
\usepackage{amsmath}
\usepackage{booktabs} 
\usepackage{soul}

\begin{document}

\title{JWST imaging of the Pleiades:  Anisotropy of turbulence in the cold neutral medium}

\authorrunning{Vigoureux et al.}

   \author{G. Vigoureux \inst{1}
          \and N. Flagey \inst{2}
          \and F. Boulanger \inst{1}
          \and A. Noriega-Crespo \inst{2}
          \and V. Guillet \inst{3}
          \and A. J. Alvarez-Castro \inst{4}
          \and N. deJesus-Rivera \inst{4}
          \and E. Allys \inst{1}
          \and J. M. Delouis \inst{5}
          \and E. Falgarone \inst{1}
          \and B. Godard \inst{1}
          \and P. Guillard \inst{6}
          \and F. Levrier \inst{1}
          \and P. Lesaffre \inst{1}
          \and A. Marcowith \inst{3}
          \and M. A. Miville-Desch\^enes  \inst{1}
          \and G. Pineau des For\^ets \inst{7,8}
          }

   \institute{Laboratoire de Physique de l'ENS, Universit\'e PSL, CNRS, F-75005 Paris   
     \and Space Telescope Science Institute, 3700 San Martin Drive, Baltimore, MD 21218
        \and Laboratoire Univers et Particules, Universit{\'e} de Montpellier, F-34095 Montpellier 
        \and Universidad de Puerto Rico, Rio Piedras, San Juan, PR 00931
         \and Laboratoire d’Océanographie Physique et Spatiale, Univ. Brest, CNRS, Ifremer, IRD, F-29200 Brest
        \and Sorbonne Universit\'e, CNRS, UMR 7095, Institut d’Astrophysique de Paris,  F-75014 Paris
        \and Universit\'e Paris-Saclay, CNRS, Institut d’Astrophysique Spatiale, F-91405, Orsay
        \and Observatoire de Paris,  Universit\'e PSL, Sorbonne Universit\'e, LUX, F-75014 Paris
        }

   \date{Received: August 27, 2025 ; Accepted: January 26, 2026 }

  \abstract 
  {Interstellar medium studies rely on magnetohydrodynamic turbulence as a framework for interpretation. In this context, the statistical characterization of interstellar observations is of prime importance. 
We present a new perspective on diffuse interstellar matter by analyzing \textit{James Webb} Space Telescope (JWST) observations of the Pleiades nebula with NIRCam. These observations are remarkable in that they provide a "microscope" view of the cold neutral medium (CNM) with a spatial resolution of 0.2\,mpc (40\,au). 
 A 2D Fourier analysis was used to characterize the structure of polycyclic aromatic hydrocarbon (PAH) emission in regions near and far from the Pleiades star Merope. To produce maps of the interstellar emission, stars and galaxies were filtered out. The final step in the data cleaning involved subtracting a component, in Fourier space, which we inferred to be a residual of the near-infrared cosmic background. 
 The PAH emission power spectra are highly anisotropic. PAH emission power spectra are highly anisotropic at all scales, with a constant level of power anisotropy. The magnetic field orientation, as derived from the {\it Planck} dust polarization data, aligns with the PAH anisotropy. The spectra are well fitted with a break-free power law, suggesting that we do not observe a specific scale for energy dissipation. Power-law indices are -3.5 near Merope and -3 in the more distant field. 
 These findings are discussed in relation to interstellar turbulence that may be driven by the Pleiades stars. The JWST observations of the Pleiades offer a new viewpoint for comparing observations and theoretical models, as they examine physical scales at which turbulence in the CNM is subsonic and decoupled from the thermal instability.  The observations may indicate that the turbulent energy cascade in the CNM is anisotropic.}


\authorrunning{Vigoureux et al.}

\titlerunning{JWST imaging of the Pleiades:  Anisotropy of the turbulent energy cascade in the CNM}
   \maketitle
%

\section{Introduction}

Studies of the cold interstellar medium (ISM) frequently rely on numerical simulations of magnetohydrodynamic (MHD) turbulence as a theoretical model \citep{Hennebelle12,Burkhart21}. In this context, the statistical analysis of interstellar observations is crucial to interpreting observations \citep{Richard25}.
This paper presents a new perspective, using observations of the Pleiades nebula by the \textit{James Webb} Space Telescope (JWST) with the Near-Infrared Camera (NIRCam).

The Pleiades reflection nebula is an iconic astronomical object familiar to a wide audience and also a prime scientific target
for studies of the dusty ISM \citep{Gibson03a,Gibson03b,Gibson07}. We are observing a chance encounter between a star cluster and diffuse molecular gas \citep{Gordon84,Herbig01,Ritchey06}. The brightest members of the Pleiades cluster are late-B stars at a distance of 135~pc \citep{Abramson18} illuminating their surrounding ISM  over a few parsecs \citep{Gibson03a}. The stellar illumination enables the ISM structure to be observed at smaller angular scales than is usually possible.
The reflection nebulosity offers a spectacular view of the filamentary structure of the diffuse ISM.

Many studies have relied on dust continuum observations to determine the power spectrum of interstellar matter \citep{Gautier92, Ingalls04, Miville07,Miville10,Miville16}. The advent of \textit{Planck} made it possible to extend this type of analysis to polarized dust emission \citep{Planck_LegacyXI}, which established the statistical alignment of diffuse interstellar structures with the magnetic field direction on parsec scales \citep{Clark15,PIPXXXII}. This result has been interpreted as a hallmark of magnetized interstellar turbulence \citep{Caldwell17,Kandel17}.
In a broader context, the JWST offers a unique opportunity. Its angular resolution and sensitivity to the emission of polycyclic aromatic hydrocarbons (PAHs) enable imaging of the cold ISM at finer scales than was previously possible. At the distance of the Pleiades, the angular resolution of NIRCam corresponds to a spatial scale of 0.2 \,mpc, at which we may be able to identify signatures of dissipation of turbulence, in particular through ion-neutral friction \citep{Burkhart15,Hu24}. Although there is observational evidence of structure in the cold neutral medium (CNM) at these scales, existing observations are too sparse to conduct a statistical study \citep{Heiles97,Stanimirovic18}. 

The purpose of this work is to demonstrate and capitalize on the potential of JWST for diffuse ISM studies. We used PAH emission to investigate the structure of the cold magnetized ISM. In fact, PAHs are  tightly coupled to the gas through collisions \citep{Omont86} and to magnetic field lines. In the physical conditions of the Pleiades nebula -- specifically the gas density and temperature  and the radiation field strength -- PAHs exist partly as cations and partly in neutral form \citep{Flagey06}. However, both charged and neutral PAHs are coupled to magnetic field lines because their charge state varies on timescales much shorter than their drag time through collisions with the gas \citep{Lepp88}. 

The paper is organized as follows. Section~\ref{sec:data} introduces the JWST observations and outlines the data processing steps to generate images of diffuse emission. Our 2D Fourier analysis of these images is described in Sects.~\ref{sec:Fourier} and~\ref{sec:spectra}. In Sect.~\ref{sec:turbulence} we discuss the results of the data analysis in relation to interstellar turbulence. Finally, Sect. \ref{sec:conclusion} summarizes our main results. The paper includes three appendices. Appendix~\ref{App:data_cleaning} details and illustrates the methods used to remove stars and mask galaxies from the JWST images. 
Appendix~\ref{App:interstellar} provides a broader context for the JWST observations, in which we characterize the magnetized interstellar matter
associated with the Pleiades nebula using {\it Planck} data. 
In Appendix ~\ref{sec:context} we use the observations to characterize the physical conditions and demonstrate that the JWST observations of the Pleiades probe the  structure of the CNM.

\begin{figure*}[ht]
    \centering
    \includegraphics[width = 500 pt]{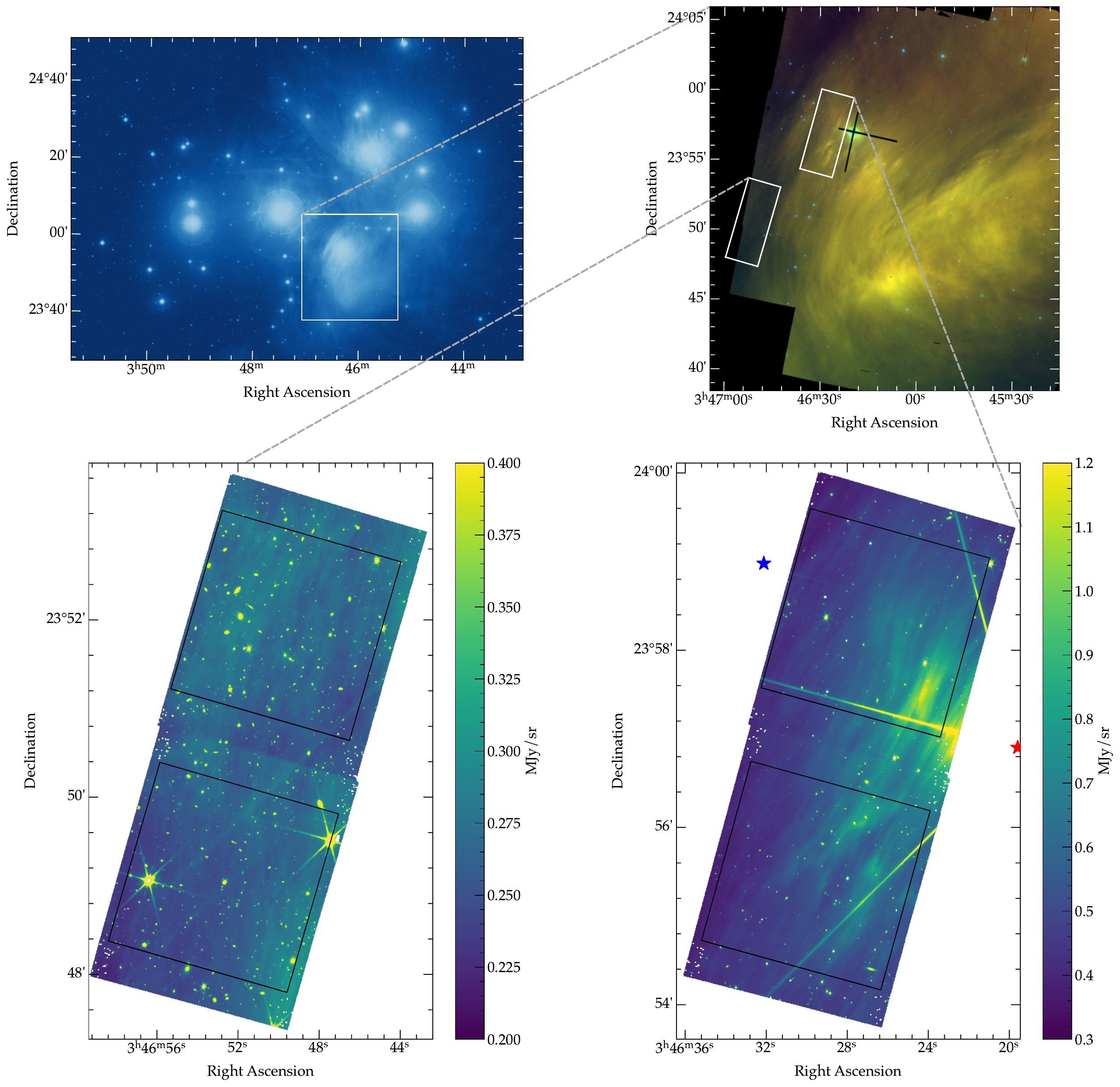}
    \caption{Overview of the F335M observations in the Pleiades. Top left: Cutout from the digitized Palomar Sky Survey (POSS~II; $\lambda_{eff}\sim480~\rm{nm}$). Top right: Three-color composition from \textit{Spitzer}/IRAC (red: 8.0~$\mu$m, green 5.8~$\mu$m, blue 3.6~$\mu$m). Merope saturates the IRAC detectors and causes the dark cross. Bottom: F335M observations at both positions in our JWST/NIRCam program. Field~2 is on the left and Field~1 on the right. Note that the color scale is not the same in the two images. The footprint of the \textit{Spitzer}/IRAC image is shown in the POSS~II image, and those of the JWST/NIRCam images are shown in the \textit{Spitzer}/IRAC image as rectangles. The red and blue stars on the rightmost JWST image indicate the positions of Merope and PQ Tau. The four black square boxes display the location of the square maps analyzed in this paper. The Cut~1 images are north of the Cut~2 images.}
\label{fig:pleiades}
\end{figure*}
  
\section{Imaging the diffuse emission from PAHs}
\label{sec:data}

This section introduces the JWST observations (Sect.~\ref{subsec:observations}) and outlines the data processing steps that were performed to derive maps of diffuse emission from PAHs (Sect.~\ref{subsec:cleaning}).

\subsection{JWST observations}
\label{subsec:observations}

The NIRCam  instrument \citep{Rieke23} was used to image two fields, each measuring $6^\prime$ by $2.2^\prime$  as part of the GO 2143 program. Field~1 is centered approximately 2\arcmin\ to the east of the bright Pleiades star Merope (23~Tau), and Field~2 about 10\arcmin\ to the southeast of Merope. The top images in Fig.~\ref{fig:pleiades} show the location of the two fields within the Pleiades. The two images at the bottom were produced by the JWST pipeline. For this paper, we used version 1.18.0 of the calibration software.

The two fields were imaged in the F212N filter, centered on the 1-0 S(1) H$_2$ line, and the F335M filter that includes the 3.3-$3.4\,\mu$m emission bands from PAHs. We also obtained observations at both sky positions in the F200W and F277W filters. The observations in the short- and long wavelength channels are executed simultaneously,  in the spectral filters F212N with F335M and F200W with F277W. 
For each pointing, the integration time was $10^4~\rm{s}$ for the H$_2$ and PAH observations, and $4000~\rm{s}$ for the wide filters. In this study, we only analyzed the F335M data. The data obtained using the three additional filters will be presented in a future paper.
  
The long channel of NIRCam comprises two square detectors, each with 2000 pixels square and a field of view of $129^{\prime\prime}$.  We used the FULLBOX 6TIGHT\footnote{NIRCam page at \url{https://jwst-docs.stsci.edu/jwst-near-infrared-camera/nircam-observing-modes/}} dithering pattern to obtain multiple offset pointings of the camera and cover the rectangular field without gaps.
The dithering also mitigates bad pixels and flat-field uncertainties; it significantly reduces the data noise on the smallest angular scales, where brightness variations are the smallest.  We did not use secondary sub-pixel dithers because the fine sampling of the point-spread function (PSF) is not essential to our observations.

\subsection{Maps of diffuse emission}
\label{subsec:cleaning}

In order to study the PAH emission,  we needed to separate the diffuse emission in the F335M filter image from numerous point sources (mostly stars) and compact sources (mostly galaxies). We did this in four main steps, which we outline here. More details are given in Appendix~\ref{App:data_cleaning}.

\begin{itemize}
\item 
We modeled and subtracted the large diffraction spikes from two bright stars, Merope and PQ~Tau. These two stars are outside the JWST fields, but their spikes are prominent structures in Field~1. One of the Merope spikes is also visible in Field~2. 

\item 
We subtracted from a first approximation the diffuse emission using a multi-scale median filter. This step yields background-subtracted images, which we then used to identify and remove point sources and compact sources located within the JWST fields. 

\item 
We subtracted point sources using an empirical template of the JWST PSF. This template was obtained by stacking background-subtracted images centered on point sources to sub-pixel accuracy. 
Point sources are subtracted by fitting the PSF template over small images centered on each source. We did not use the core of the point sources in this fit because it is often saturated. 

\item 
Compact extended sources — including galaxies and the saturated cores of bright stars — are identified on maps after background, spike, and point source subtraction, and are then masked. 
The sky area we masked is set by a brightness threshold applied to maps in which spikes, multi-scale background, and point sources have  been subtracted (see Appendix~\ref{App:data_cleaning}). The masked areas are filled in with the multi-scale background maps.

\end{itemize}

To assess systematic uncertainties associated with the data cleaning, we  checked how sensitive the results of our data analysis are to the brightness threshold used to create the mask. For the fiducial mask, 9\% of the sky pixels are masked. We used two alternative masks, the light mask (for which this fraction is 5\%) and the heavy mask (14\%).

For statistical analysis, we divided the dither pointings into two sets $h1$ and $h2$, which we used to produce a pair of images for each of the two fields. The two images in each pair cover the same sky region, but have independent statistical noise. Each of the four images (2 pairs) is processed separately by following steps (1) to (4). Separate background maps and masks are computed for each image. After step (4), the difference between the two images in each pair allows us to identify and mask a small number of glitches present in one of the two images.  
Mathematically, for each field, we obtained two independent maps of diffuse emission, $d_{h1}$ and $d_{h2}$, from  sky images $s_{h1}$ and $s_{h2}$ with spikes and stars subtracted:  
\begin{eqnarray}
\begin{aligned}
\label{eq:im_painting}
&d_{h1} = M_{h1} \, s_{h1} + (1-M_{h1}) \, b_{h1} \\
&d_{h2} = M_{h2} \, s_{h2} + (1-M_{h2}) \, b_{h2}, 
\end{aligned}
\end{eqnarray}
where $M_{h1}$ and $M_{h2}$ denote masks and $b_{h1}$ and $b_{h2}$ the background maps. 

\begin{figure*}[ht]
\centering
\includegraphics[width = 500 pt]{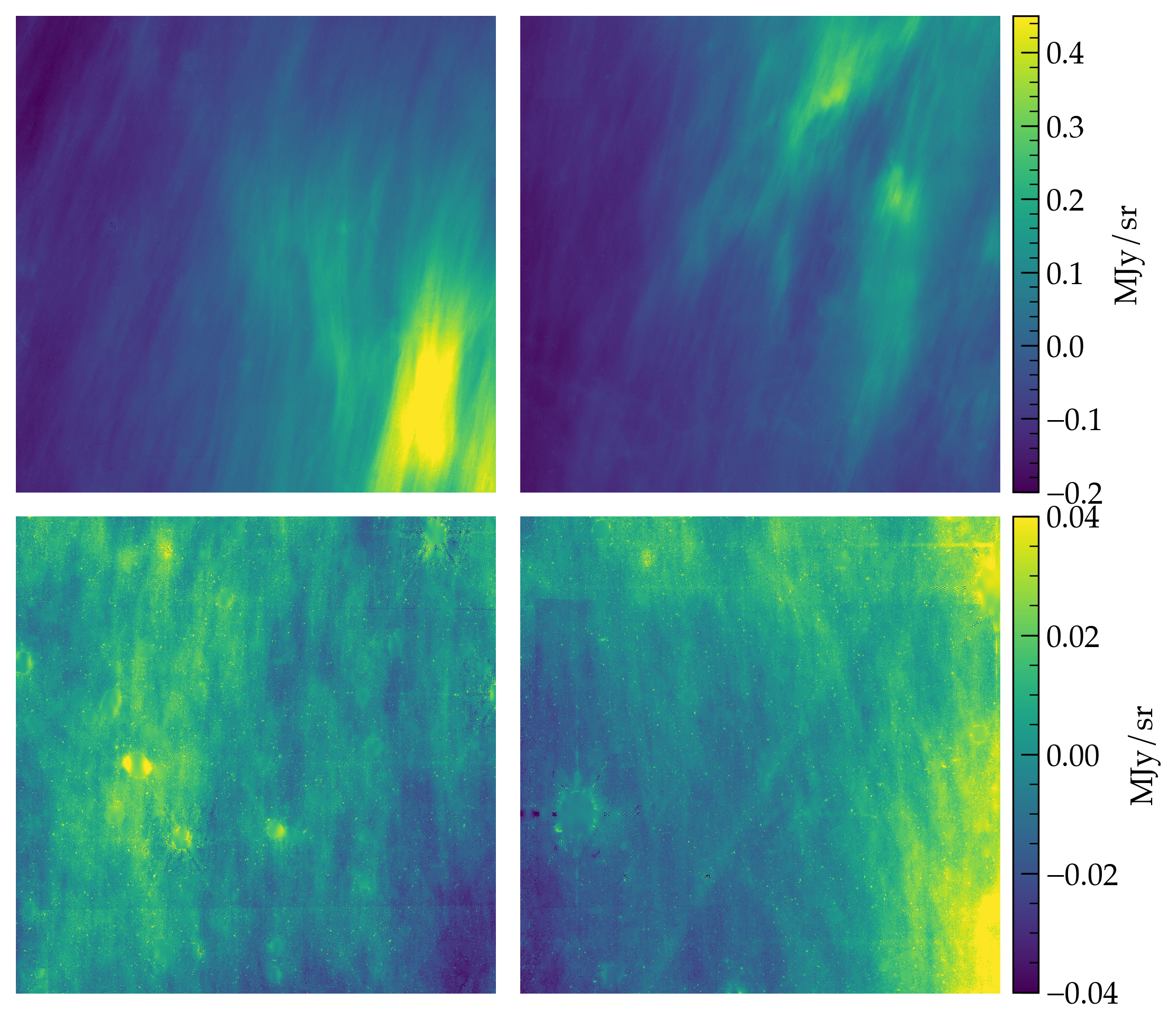}
\caption{ JWST images of diffuse sky emission. The panels show the average image of the four pairs we analyzed; the top row shows Field~1 images, and the bottom row Field~2 images. The Cut~1 images are to the left and the Cut~2 images to the right. The mean values of each image are set to zero.  The center positions and the position angles of the vertical axes of the images can be found in Table~\ref{tab:coordinates}. }   
\label{fig:four_cleaned_images}
\end{figure*}

For power spectrum analysis, we cut out two square images, with 2000 pixels on each side, in each field. The positions of these cuts are drawn with solid black lines in the JWST images displayed in Fig.~\ref{fig:pleiades}. The center positions and orientation of these images can be found in Table~\ref{tab:coordinates}. We analyzed eight square images (four pairs) from which diffraction spikes, stars, galaxies, and glitches have been removed.
The maps of diffuse emission in each of the four pairs are presented in Fig.~\ref{fig:four_cleaned_images}.  The four images show a similar pattern with nearly parallel striations. The JWST/NIRCam pixel size $\mathrm{pix_{JWST}} = 0.063^{\prime\prime}$ corresponds to $4.1\,10^{-2}\,$mpc (8.5\,au) at the distance of the Pleiades. The full size of each square image is 80\,mpc. Compared with other observational means, one can say that the JWST offers us a "microscope" view at the diffuse ISM. 

\begin{figure}[ht]
\centering
\includegraphics[width = 250 pt]{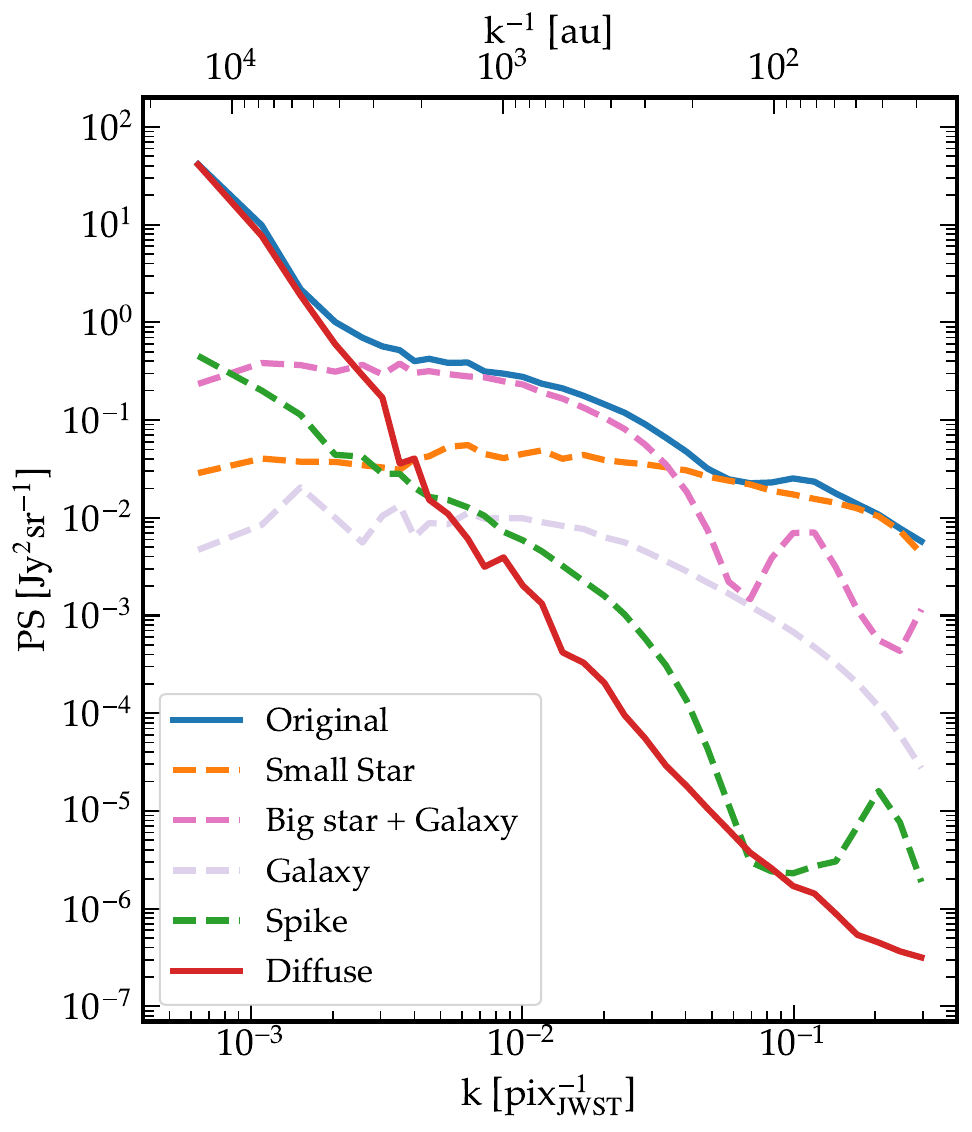}
\caption{Power spectra of Field~1 Cut~1 at different steps of the data cleaning. 
The top spectrum is that of the initial JWST image with the stars, their spikes, and galaxies. The bottom is that of the image at the end of cleaning. The dashed lines  show the spectra of the maps of stars, spikes, and galaxies subtracted from the JWST image. The wavenumber is expressed in units of $\rm pix^{-1}_{JWST}$ on the bottom axis and au on the top axis.  }
\label{fig:power_spectra_evolution}
\end{figure}

Pairs of square images are used to compute cross-power spectra unbiased by data noise.
To emphasize and demonstrate the importance of data cleaning, we calculated the cross-power spectra at each stage for Field~1 Cut~1 as an example. The results of this analysis are shown in Fig.~\ref{fig:power_spectra_evolution}. The top spectrum represents the unprocessed JWST image that features spikes, stars,  and galaxies. In contrast, the bottom spectrum corresponds to the final image of the diffuse emission. The intermediate curves indicate the cross-power spectra of the maps of spikes, stars, and galaxies that were subtracted from the original data. Data cleaning leads to a decrease in the amplitude of the spectra by up to 4 orders of magnitude for the highest $k$. 

The mean brightness has been subtracted from the JWST images because it is a model value \citep{Rigby23} introduced in the data pipeline that does not separate interstellar emission from the zodiacal, stellar, and extragalactic backgrounds. 
To estimate the mean PAH emission, we measured the mean brightness averaged over each image field of view at $12\,\mu$m, using the Wide-Field Infrared Explorer (WISE)  data \citep{Wright10}. The values listed in Table~\ref{tab:coordinates} represent the emission from the Pleiades nebula with the large-scale Galactic contribution, measured off the nebula, subtracted out.  The brightness ratios between the diffuse Galactic emission in the F335M NIRCam filter and the WISE $12\,\mu$m filter, $I_{3.5}/I_{12}$ values listed in Table~\ref{tab:coordinates}, were measured by correlating the structure of the 12$\,\mu$m  emission with the multi-scale background F335M maps. These values are consistent with dust emission models \citep{Compiegne11,Draine07} and observations\footnote{The color ratio $I_{3.5}/I_{12}$ is related to the size distribution of PAHs. The diffuse ISM values, derived from the analysis of the DIRBE and \textit{Spitzer}/IRAC data, are in the range 2 to $6\,10^{-2}$ \citep{Dwek97,Flagey06}}.

\begin{table*}[ht]
    \caption{JWST cutout images.}
    \label{tab:coordinates}
    \begin{tabular}{lccccc}
    \hline 

    Image   & RA & Dec & PA & $I_{\mathrm{12}}$ & $10^2 \times I_{3.5}/I_{12}$ \\
    & J2000  & J2000  & deg & MJy\,sr$^{-1}$ & \\ 
    & (a) & (a) & (b) & (c) & (d) \\
    \hline 
    {\bf Field1}  &  & & -15.4  & & \\ 
    Cut~1  &  3:46:26.61 & 23:58:18.5 & & $7.0\pm 0.2$  & $11.6 \pm 0.7 $ \\ 
    Cut~2  &  3:46:29.55  & 23:55:27.4 & & $8.5\pm 0.2$ & $ 6.1 \pm 1.0$ \\ 
    {\bf Field2}  &  & & -16.1  &  \\ 
    Cut~1  &  3:46:49.64 & 23:51:56.3 & & $3.7 \pm 0.2$ & $4.4 \pm 1.8 $\\ 
    Cut~2  & 3:46:52.72  & 23:49:5.7 & & $3.3 \pm 0.2$ & $3.5 \pm 0.9$\\ 
    \hline
    \end{tabular}

\tablefoot{(a) Coordinates of image centers. (b) Position angle of the long axis of the JWST mosaic with respect to the north direction measured positive toward the east (counterclockwise direction). (c) Mean emission at $12\, \mu$m averaged over each of the Pleiades images, measured using the WISE full resolution image. The background contribution is subtracted. (d) Emission ratio derived from a linear correlation of the NIRCam F335M background image and the WISE data.} 
\end{table*}

\begin{figure}[ht]
\centering
\includegraphics[width = 240 pt]{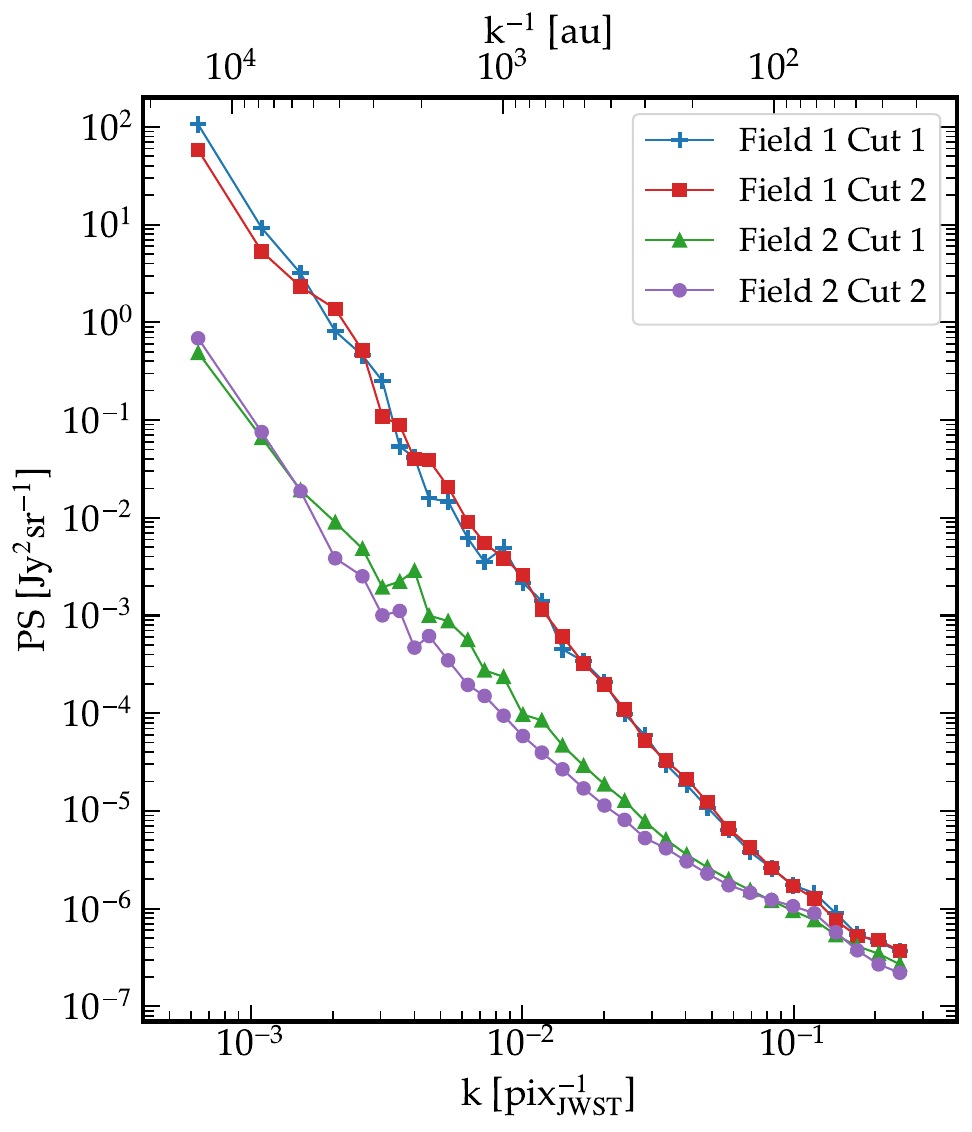}
\caption{Cross-power spectra for the four JWST images.  Distinct colors and symbols identify the spectra of the images listed in the top-right corner. The spectra are corrected for beam attenuation. Due to the large dynamic range of the four spectra, the statistical error bars are too small to be visible on the plot.   }   
\label{fig:four_JWST_spectra}
\end{figure}

\section{Fourier analysis}
\label{sec:Fourier}

This section presents the Fourier analysis of the images of diffuse emission shown in Fig.~\ref{fig:four_cleaned_images}. 

\subsection{Computation and statistical uncertainties}

We computed the 2D cross-power spectrum, $P_{2D}(k,\theta),$ for the pair of square images $d_{h1}$ and $d_{h2}$ introduced in Eq. \ref{eq:im_painting}:
\begin{equation}
P_{2D}(k,\theta) = \Re \,\left(\widehat{M_A \, d_{h1}} \times \widehat{M_A \, d_{h2}}^\ast \right),
\label{Eq:PS2D}
\end{equation}
where the wavenumber $k$ and the angle $\theta$, measured from the celestial north modulo $\pi$, are polar coordinates in the Fourier plane. $\Re $ denotes the real part, the hat symbol the Fourier transform, and the asterisk the complex conjugate. 
$M_A$ represents the image apodization applied to avoid edge effects in the Fourier transform\footnote{We used the Tukey function in the SciPy python package with its default parameters.}.  

We use $P_b^{d_1\times d_2}(k,\theta)$ to denote the mean value of $P_{2D}(k,\theta)$, within bins of width $\Delta k$ and $\Delta \theta$, and $P_b(k)$ the cross-power spectrum averaged over angles. 
To estimate uncertainties on  $P_b^{\, d_1\times d_2}(k,\theta)$ and $P_b(k)$, we used the following formulae:
\begin{align} 
& \sigma \, (P^{\, d_{h1}\times d_{h2}}_b)\,^2 = \left(2\,(P^{\, d_{h1}\times d_{h2}}_b)^2+2\,P^{\, d_{h1}\times d_{h2}}_b \times N + N^2\right)/n_b \label{eq:PS_uncertainties} \\
& n_b = \frac{4\,\pi}{A_\mathrm{map}} \, \Delta \theta \, k \,\Delta k, \label{eq:nb} 
\end{align} 
where $N$ is the noise power spectrum of images $d_{h1}$ and $d_{h2}$, $n_b$ is the number of Fourier modes in the bin $b$, and $A_\mathrm{map}$ is the effective area of the sky covered by the maps after apodization. We refer to \citet{Knox95} and \citet{Hivon02} for a mathematical derivation and the numerical assessment of these formulae, and to \citet{Viero13} and \citet{Cordova24} for their application to the analysis of infrared sky maps. To apply this formula, we required the noise power spectra $N$ for each pair of square maps.  These were estimated from the power spectrum of the difference $d_{h1}-d_{h2}$, for Fields 1 and 2 separately.  The factor 2 arising from the difference was taken into account.   

The left term in Eq.~\ref{eq:PS_uncertainties} represents the cosmic variance, the minimum uncertainty on the sky spectrum estimate for noise-free data, the right term is the variance of the data noise, and the middle term the random correlation between sky emission and data noise. The left term dominates for the lowest wavenumbers, where the maps comprise a small number of independent modes, and the right term dominates for the highest wavenumbers, where the signal-to-noise ratio on the sky emission is the lowest.

Figure~\ref{fig:four_JWST_spectra} shows the cross-power spectra $P_b(k)$ of the four JWST images. Hereafter, the wavenumber expressed in units of $\rm pix^{-1}_{JWST}$ in the bottom axis is converted into physical scales in au on the top axis for a distance to the Pleiades of 135\,pc \citep{Melis14,Abramson18}. 
The spectra are corrected for the signal attenuation by the telescope beam, using the power spectrum of the PSF image produced with the STPSF Python package for our NIRCam observations in the F335M filter. 
The PSF varies across the field of the camera, but this effect is ignored because it is smeared out by the raster mapping of the sky. 
The spectra in Fig.~\ref{fig:four_JWST_spectra}  all show a flattening at high $k$. This is not due to data noise, because the noise is not correlated between maps $d_{h1}$ and $d_{h2}$.

\begin{figure}[ht]
\centering
\includegraphics[width = 240 pt]{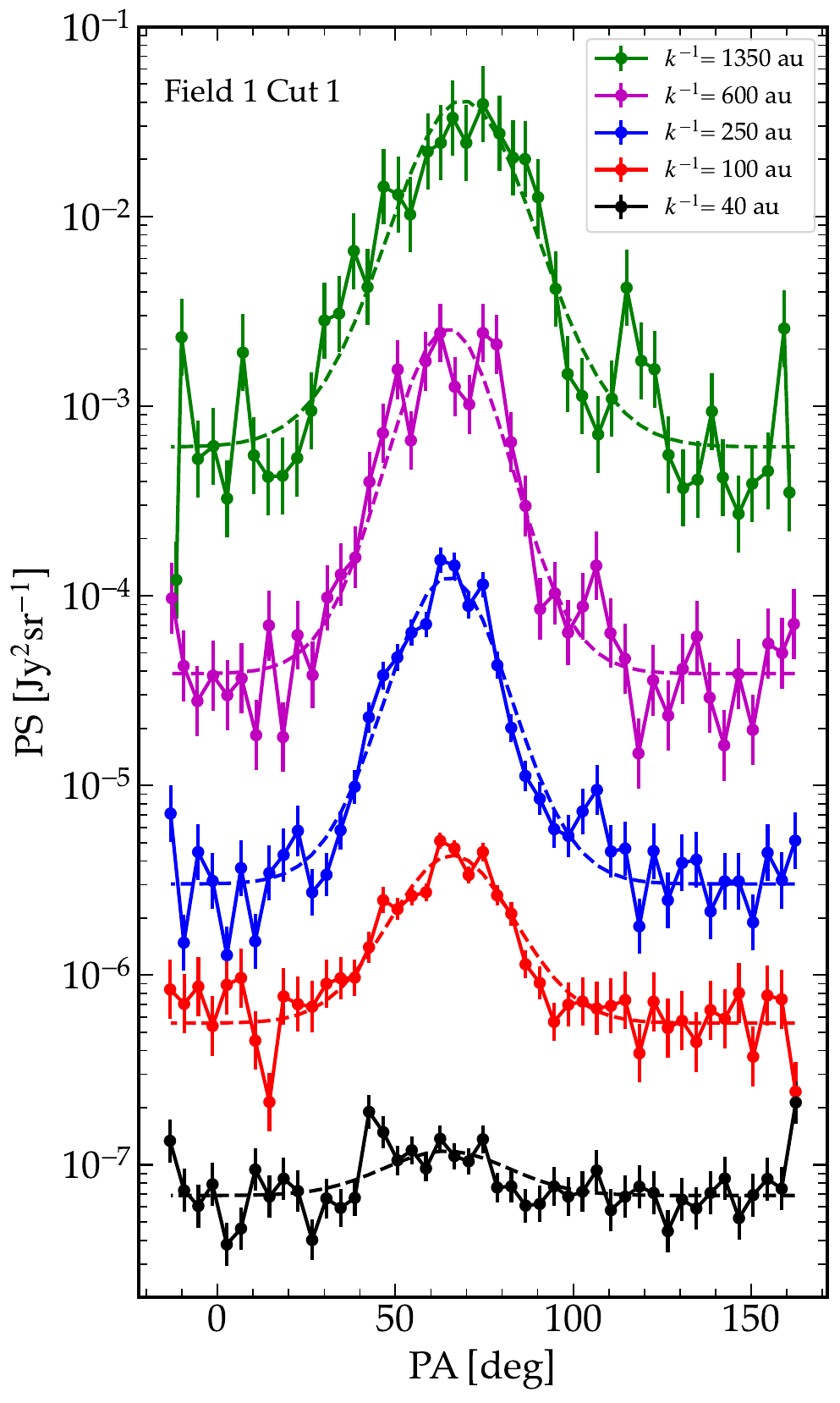}
\caption{Angular dependence of the power spectrum amplitude for the Field~1 image Cut~1. The angular dependence  is plotted for the 5 $k$ values listed in the top-right corner. The position angle (horizontal axis) is measured modulo $\pi$ with respect to the celestial north and positive to the east. The dashed lines show the fits of the Gaussian model in Eq.~\ref{eq:PS_model} to the profiles.  }   
\label{fig:angular_profiles}
\end{figure}

\subsection{Anisotropy model}
\label{subsec:aniso_model}

To analyze the anisotropy of the diffuse emission, we averaged $P_{2D}(k, \theta) $ into logarithmic bins of k with $\Delta k/k = 0.2 $ and linear bins of $\theta $, modulo $\pi$, with  $\Delta \theta = 4^\circ$, which do not overlap. 
We use $k_i$ and $\theta_j$ to denote the bin centers and $P_{i,j}$ the binned value of $P_{2D}$. The uncertainty on $P_{i,j}$ is estimated using Eq.~\ref{eq:PS_uncertainties}.
We modeled the angular dependence of $\mathrm{log} \, P_{i,j}$ with a Gaussian profile. Mathematically our data model is as follows:
\begin{equation}
\mathrm{log}\, (P_{i,j}) =   \Gamma_i \, {\rm exp}\,\left(-\frac{(\theta_j - \theta_0)^2}{2 \, \sigma_\theta^2}\right) + B_i.
\label{eq:PS_model}
\end{equation}
The angle $\theta_0$ defines the orientation, perpendicular to the striations seen in the images, for which the power is maximum, and $\sigma_\theta$ measures the dispersion of this orientation over the images. $\Gamma_i$ represents the ratio of maximum to minimum power while  $B_i$  the minimum power within the $k_i$ bin. We performed a global fit of $\mathrm{log}(P_{i,j})$ with values of $\theta_0$ and $\sigma_\theta$ independent of $k$. These two model parameters are listed in Table~\ref{tab:angles} for each of the four images in Fig.~\ref{fig:four_cleaned_images}. The values of $\theta_0$ match the dust polarization angles measured at the image centers using \textit{Planck} polarization data at 353\,GHz (see Appendix~\ref{App:interstellar} and Fig.~\ref{fig:Pleiades_Planck_polar}). In emission, the dust polarization angle is perpendicular to the projection of the magnetic field on the plane of the sky. As $\theta_0$ measures the orientation perpendicular to the striations, we conclude that the filamentary structure of the PAH emission is aligned with the magnetic field projection. This is  similar to what is observed on parsec scales with {\it Planck} data \citep{Clark15,PIPXXXII}.

Figure~\ref{fig:angular_profiles} presents the angular profiles of the power spectrum amplitude obtained for Cut~1 in Field~1. The profiles are shown for five $k$-bins, equally spaced in log($k$) from large to small angular scales from top to bottom; the dashed lines represent the model fit to the data.  
We note that the profile appears to be less anisotropic for high wavenumbers\footnote{ The fit shown in Fig.~\ref{fig:angular_profiles} allows for small variations of $\theta_0$ and $\sigma_\theta$ between the five $k$ bins.}. 

The model allowed us to derive the power spectra for the orientations parallel and perpendicular to the striations in the sky image:
\begin{eqnarray}
\begin{aligned}
&P_{i,\parallel} \equiv 10^{B_i}\, / \, P_\mathrm{PSF} \\
&P_{i,\perp} \equiv 10^{(\, \Gamma_i+B_i \,)}  \, / \, P_\mathrm{PSF},
\label{eq:power_spectra1}
\end{aligned}
\end{eqnarray}
where $P_\mathrm{PSF}$ is the power spectrum of the JWST PSF. The division by $P_\mathrm{PSF}$ corrects the spectra for the attenuation of the beam. The data error bars are propagated to $P_\parallel$ and $P_\perp$ by the fitting procedure. Our definition of $P_{\parallel}$ corresponds to the power spectrum of the emission parallel to the mean orientation of the striations\footnote{This is true provided that the Gaussian in Eq.~\ref{eq:PS_model} is negligible at $90^\circ$ of $\theta_0$, a condition that is satisfied for the values of $\sigma_\theta$ in Table~\ref{tab:angles}.}. As \citet{Cho00} pointed out in their analysis of MHD turbulence simulations, the Fourier transformation only partially  measures the 3D anisotropy of the PAH emission because the striations orientation varies across the sky and along the line of sight. The angular widths of the profiles in Fig.~\ref{fig:angular_profiles} reflect the dispersion of the orientation of the striations across the image. These widths are significant. They exceed the dispersion of the polarization angles ($12^\circ$) reported by \citet{Bracco16} in their analysis of {\it Planck} data in the southern Galactic cap.

\begin{table}[ht]
    \caption{Parameters of the anisotropy fit and dust polarization angles.}
    \label{tab:angles}
    \begin{tabular}{lccc}
    \hline 

    Field   & $\theta_0$ & $\sigma_\theta $ & $\psi$\\
    & deg.  & deg. & deg. \\
    & (a) & (a) & (b) \\ 
    \hline 
    Field~1  &  &  & \\ 
    Cut~1  &  $64.9 \pm  0.2$ & $17.7  \pm 0.2$  & $57 \pm 11$\\ 
    Cut~2  &  $58.6 \pm  0.2$  &  $15.4  \pm 0.2$  & $62 \pm 7$\\ 
    Field~2  &  & &    \\ 
    Cut~1  &  $72.6 \pm  0.6$ & $20.2  \pm 0.8$ & $77 \pm 9$ \\ 
    Cut~2  &  $81.6 \pm  0.8$  &  $ 20.8  \pm 1.3 $ & $86 \pm 5$\\ 
    \hline
    \end{tabular}

\tablefoot{(a) Parameters characterizing the 
angle dependence of $P_{2D}$ in Eq.~\ref{eq:PS_model}. (b) Dust polarization angle at the position of the image centers measured using {\it Planck} polarization data at 353\,GHz (see Appendix~\ref{App:interstellar}).}
\end{table}

\subsection{Cosmic near-infrared background residual}
\label{subsec:residual}

Figures~\ref{fig:four_JWST_spectra} and \ref{fig:angular_profiles} indicate that the power spectra include an isotropic emission component, which prevails at high $k$ values. We characterized this component, which we identified as an extragalactic residual left after data cleaning.

\begin{table}[ht]
    \caption{Power-law fits of the residual  spectrum. }
    \label{tab:PSfit_residual}
    \begin{tabular}{lccc}
        \hline 
\\[-1.0ex]

    Mask   & $A_\mathrm{res}$  & $\alpha_\mathrm{res}$ & $\chi^2$  \\
& [Jy$^2$\,sr$^{-1}$] & & \\
            & (a)     & (b) & (c)  \\
            \\[-1.0ex]
   \hline 
   \\[-1.0ex]
    Fiducial &  $1.4 \pm 0.1 \, 10^{-5}$   &  $-1.23 \pm 0.025$ &2.8 \\ 
     Light  & $5.9 \pm 0.8 \, 10^{-5}$   & $-1.40 \pm 0.056$  &24 \\ 
     Heavy  & $1.2 \pm 0.2 \, 10^{-5}$   & $-1.24 \pm 0.060$ & 13   \\  
        \hline
    \end{tabular}
\\[1.0ex]
\tablefoot{(a): Amplitude of the power law at wavenumber $k_0 = 10^{-2}\,$pix$_\mathrm{JWST}^{-1}$. 
(b): Spectral index of the power law. 
(c): Reduced $\chi^2$ of the fit.}
\end{table}

\begin{figure}[ht]
\centering
\includegraphics[width = 240 pt]{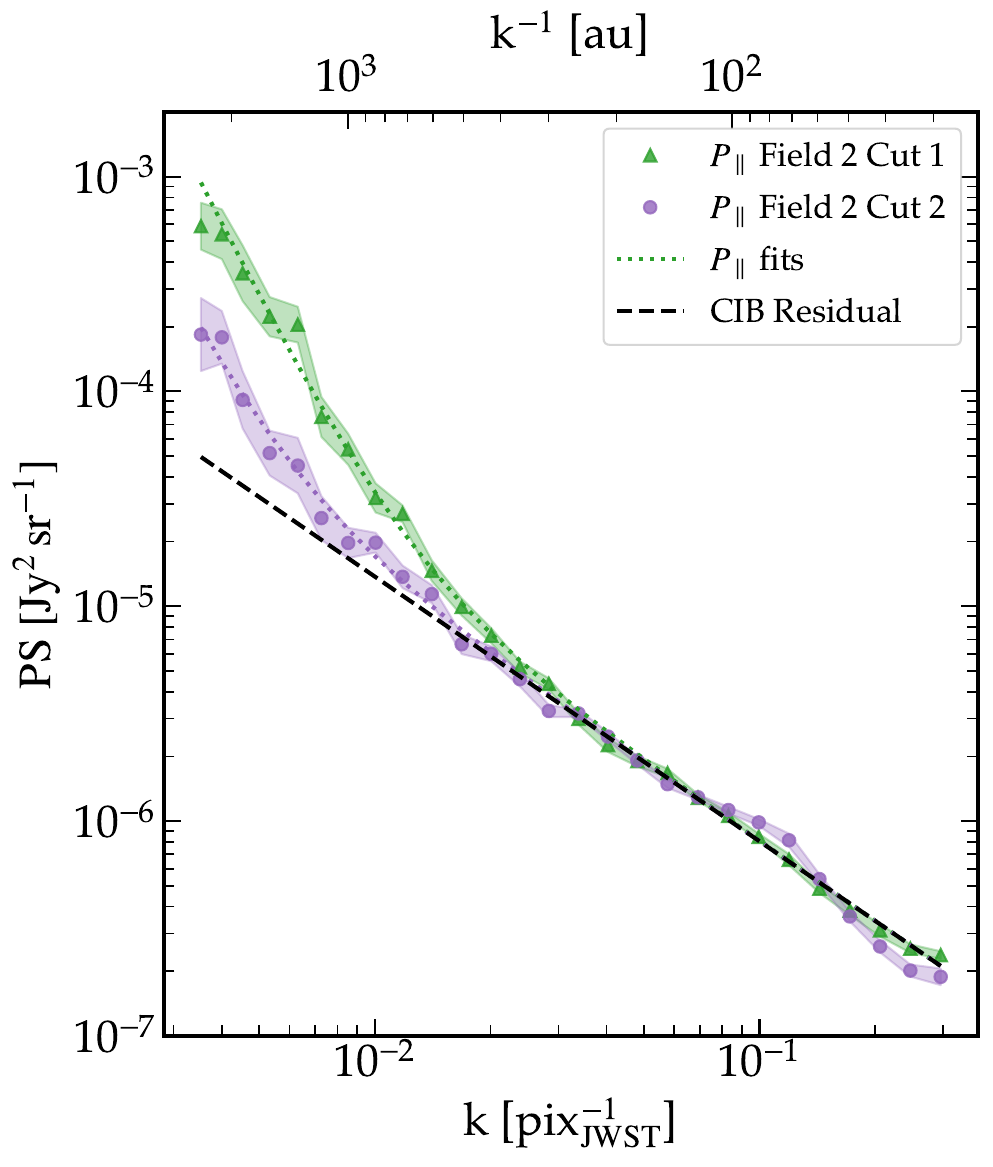}
\caption{Power spectra for the parallel orientation of Field~2 Cuts~1 and 2. The color shadings indicate $1\sigma$ uncertainties. The dashed black line shows the power-law fit of the  component that we interpret as a CIB residual. The dotted lines show the data fit using the sum of two power laws to account for the PAH and residual components. }
\label{fig:residual}
\end{figure}

The isotropic component is most readily apparent in the $P_\parallel$ spectra of Field 2 Cuts~1 and 2 displayed in Fig.~\ref{fig:residual}, because in these spectra the PAH contribution is the smallest. One can clearly see that the two spectra flatten toward high $k$. 
Figure~\ref{fig:residual} shows fits of the $P_\parallel$ spectra with a model including two power laws: a steep one representing the PAH contribution and a shallow one representing the isotropic component. For the isotropic component, the same power law is used for Cuts~1 and 2. We performed the data fit for the fiducial, light and heavy masks. The amplitudes and spectral indices of the isotropic component are listed in Table~\ref{tab:PSfit_residual}. The amplitude and spectral index of the isotropic component are consistent for the fiducial and heavy masks.  The reduced $\chi^2$ of the fit is good for the fiducial mask but poor for the light and heavy masks. For the light mask, the isotropic component is dominant for all $k$ and the fit does not succeed in separating the PAH and the isotropic components. For the heavy mask, the amplitude of the isotropic component differs significantly between Cuts~1 and 2. 

The spectral index of the isotropic component, about $-1.2$, coincides with the value measured for  the clustering of faint extragalactic sources combined into a diffuse extragalactic background. 
The cosmic near-infrared background (CIB)  was detected with Infrared Array Camera (IRAC) \textit{Spitzer} data. The measurements reported by \citet{Kashlinsky12} for the IRAC $3.5\,\mu$m channel can be directly compared with the results of our JWST analysis. The spectral index of the isotropic component is consistent with the \textit{Spitzer} CIB power spectra, and the amplitude is one order of magnitude smaller. We conclude that the isotropic component is a residual CIB, which remains after the masking of extragalactic sources\footnote{We note that
the  spectra in Fig.~\ref{fig:residual} do not exhibit flattening at high $k$, which could be interpreted as shot noise from faint sources at the JWST detection limit.}.

\begin{figure}[ht]
\centering
\includegraphics[width = 249 pt]{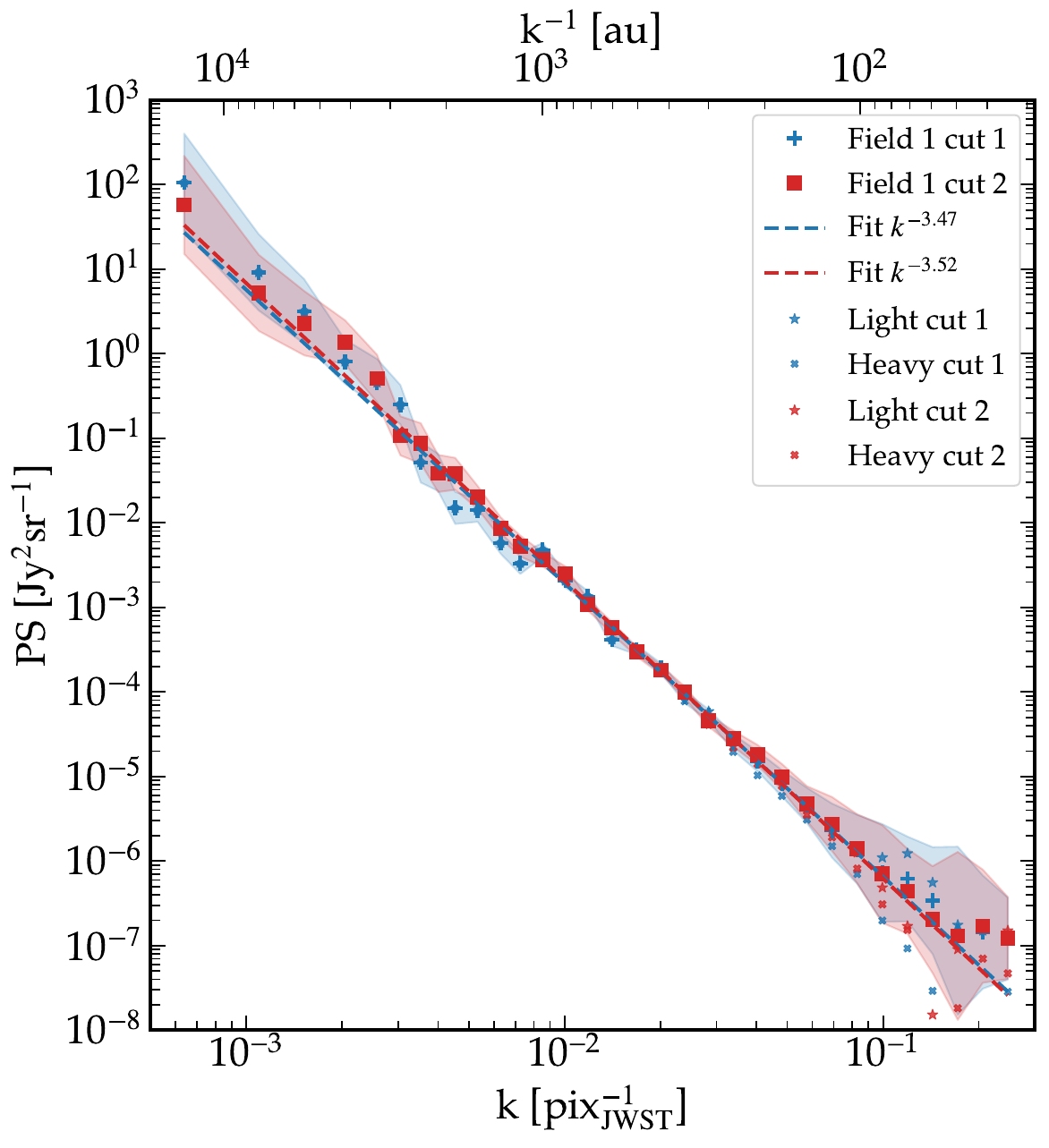}
\caption{Power spectra computed for JWST Field 1 images Cuts~1 and 2. The color shadings display $1\sigma$ uncertainties. The dashed lines show the power-law fits. The spectra obtained with the light and heavy masks are plotted with smaller symbols: circles and stars.   }
\label{fig:PS1d}
\end{figure}

\begin{table}[ht]
    \caption{Power-law fits of the difference between the spectra of Field~1 and Field~2.}
    \label{tab:PS1d}
    \begin{tabular}{lcccc}
        \hline 
\\[-1.0ex]

    {\bf Image}   & $10^3\,A_d$  & $\alpha_d$ & Mask & $I_{3.5}$ \\
& [Jy$^2$\,sr$^{-1}$] & & & [MJy\,sr$^{-1}$] \\
            & (a)     & (b) & (c) & (d) \\
            \\[-1.0ex]
   \hline 
   \\[-1.0ex]
    Cut~1  &  $1.93 \pm 0.14 $   &  $-3.47 \pm 0.07$ & F & $0.26$    \\ 
       & $1.95 \pm 0.14 $   & $-3.43 \pm 0.07$ & L &     \\ 
       & $1.91 \pm 0.15 $   & $-3.66 \pm 0.07$ & H &    \\  
    Cut~2  &  $2.06 \pm 0.09 $  &  $-3.52 \pm 0.04$  &F & $0.25$  \\ 
     &  $2.09 \pm 0.11 $  &  $-3.53 \pm 0.05$  &L &   \\ 
     &  $2.03 \pm 0.10 $  &  $-3.65 \pm 0.05$  &H  &   \\ 
        \hline
    \end{tabular}
\\[1.0ex]
\tablefoot{(a): Amplitude of the power law at wavenumber $k_0 = 10^{-2}\,$pix$_\mathrm{JWST}^{-1}$. 
(b): Spectral index of the power law. 
(c): Mask used to clean the data. F, L and H stand for fiducial, light and heavy masks (see Sect.~\ref{subsec:cleaning}).
(d): Median brightness at $3.5\,\mu$m of diffuse emission computed on the Field~1 - Field~2 difference maps.} 
\end{table}

\section{Power spectra of PAH emission }
\label{sec:spectra}

In this section we derive power spectra of the Pleiades PAH emission using JWST data. To obtain these spectra, we needed to subtract the CIB residual. This was achieved through three distinct methods, each based on different hypotheses. 
In Sect.~\ref{subsec:ON-OFF} we assume that the CIB residual is the same in our two fields, both close to and far from Merope, and we used the latter for subtraction. In Sect.~\ref{subsec:aniso} we use the anisotropy of the Pleiades emission to subtract the isotropic CIB residual. Section~\ref{subsec:mcmc} presents fits of the Field~1 spectra using two power laws to represent the PAH contribution and the CIB residual.

\begin{table}[ht]
    \caption{Power-law fits of spectra from the anisotropy difference.}
    \label{tab:PS_aniso}
    \begin{tabular}{lccc}
    \hline
    Field    & $A_a$ & $ \alpha_a$  & Mask \\
     & $\mathrm{Jy^2\,sr^{-1}}$ & \\ 
     & \multicolumn{2}{c}{(a)} & (b) \\
    \hline 
    Field~1   &  & &  \\ 
    Cut~1  & $1.3 \pm 0.1\, 10^{-2}$ & $-3.54 \pm 0.05 $ & F \\ 
    & $1.3 \pm 0.1\, 10^{-2}$ & $-3.53 \pm 0.05 $ & L \\
    & $1.3 \pm 0.1\, 10^{-2}$ & $-3.63 \pm 0.06 $ & H \\
    Cut~2  & $1.3 \pm 0.1\, 10^{-2}$ & $-3.49 \pm 0.04 $ & F \\ 
& $1.2 \pm 0.1\, 10^{-2}$ & $-3.44 \pm 0.04 $ & L \\    
& $1.3 \pm 0.1\, 10^{-2}$ & $-3.56 \pm 0.05 $ & H \\
\hline
    Field~2   &  & & \\ 
    Cut~1  & $3.5\pm 0.3 \,10^{-4} $  & $-3.05 \pm 0.06$  & F \\ 
     & $2.6\pm 0.3 \,10^{-4} $  & $-2.82 \pm 0.08$  & L \\ 
    & $3.8\pm 0.4 \,10^{-4} $  & $-3.15 \pm 0.06$  & H \\ 
    Cut~2 & $1.4\pm 0.1 \,10^{-4} $ & $-2.94 \pm 0.06$  & F \\ 
    & $1.4 \pm 0.1 \,10^{-4} $  & $-2.80 \pm 0.06$  & L \\    
   & $1.4 \pm 0.1 \,10^{-4} $  & $-2.94 \pm 0.06$  & H \\ 
    \hline
    \end{tabular}

\tablefoot{(a): Parameters of the power-law fit to $P_\perp - P_\parallel $ for $k_0 = 10^{-2} \, \mathrm{pix_{JWST}^{-1}}$ in Eq.~\ref{Eq:pwfit_dif}. 
(b): Masks used to compute power spectra. F, L and H stand for fiducial, light and heavy.}
\end{table}

\subsection{Power spectra from the spatial difference}
\label{subsec:ON-OFF}

For each spectrum $P_b(k)$ obtained for Cuts~1 and 2 of Field~1, we subtracted the mean of the Field~2 power spectra for Cuts~1 and 2.
The difference subtracts the CIB residual and a potential contribution from the Galactic background to the PAH spectra. This approach does not make use of the anisotropy fit, which restricts the range of $k$ bins that can be measured with relevant accuracy. The differences in spectra
are presented in Fig.~\ref{fig:PS1d}. 
The dashed lines represent power-law fits: 
\begin{equation}
\Delta P_b(k)/P_\mathrm{PSF}(k) =  A_d \left(\frac{k}{k_0}\right)^{\alpha_d},
\label{Eq:pwfit}
\end{equation} 
where the division by $P_\mathrm{PSF}$ corrects for the signal attenuation by the JWST beam.

The fit parameters, the amplitude $A_d$ and the spectral index $\alpha_d$, are listed in Table~\ref{tab:PS1d}. The table also includes the mean $\bar{I}$ of the sky emission measured after source masking and OFF subtraction. Figure~\ref{fig:PS1d} shows the spectra, and Table~\ref{tab:PS1d} lists the fit results, for the three masks. The light and heavy masks are used to assess systematic uncertainties. The data points in Fig.~\ref{fig:PS1d} show a noticeable steepening of the spectra at high $k$ for the heavy mask. The difference in the spectral index is slightly larger than the statistical error.  This may be due to the lack of small-scale structure in the multi-scale background used to fill in the masked areas. 

In this work, we assumed that on the spatial scales probed by the Pleiades JWST data, PAH emission traces the gas column density. For scales smaller than the line-of-sight depth of the emitting medium, the spectral index of the column density power spectrum is identical to that of the 3D density field \citep{Cho03,Brunt10}. We therefore proceed under the assumption that the PAH power spectrum is a  proxy for the gas density power spectrum. 

The spectral index that we measure matches the reference spectral index for MHD turbulence \citep{Iroshnikov64,Kraichnan65}. It is also consistent with the index inferred by \citet{Armstrong95} for electron density fluctuations in the local ISM by analyzing pulsar scintillation measurements on scales below $0.3\,$mpc.

To compare with previous measurements of dust power spectra of the diffuse ISM, we refer to the spectra measured by \citet{Gibson07} in the Pleiades and by \citet{Miville16} for an independent field at high Galactic latitude. These spectra, which were obtained by analyzing images of dust scattered light, are the closest observational match to our analysis in terms of angular resolution. For far-infrared dust emission, we also consider results derived from \textit{Herschel} observations by \citet{Miville10} and \citet{Auclair24} with angular resolutions of 30 and $15^{\prime\prime}$, respectively. The spectra reported in these four dust studies with spectral indices ranging from -2.7 to -2.95 are shallower than the two shown in Fig.~\ref{fig:PS1d} .

\begin{figure}[ht]
\centering
\includegraphics[width = 245 pt]{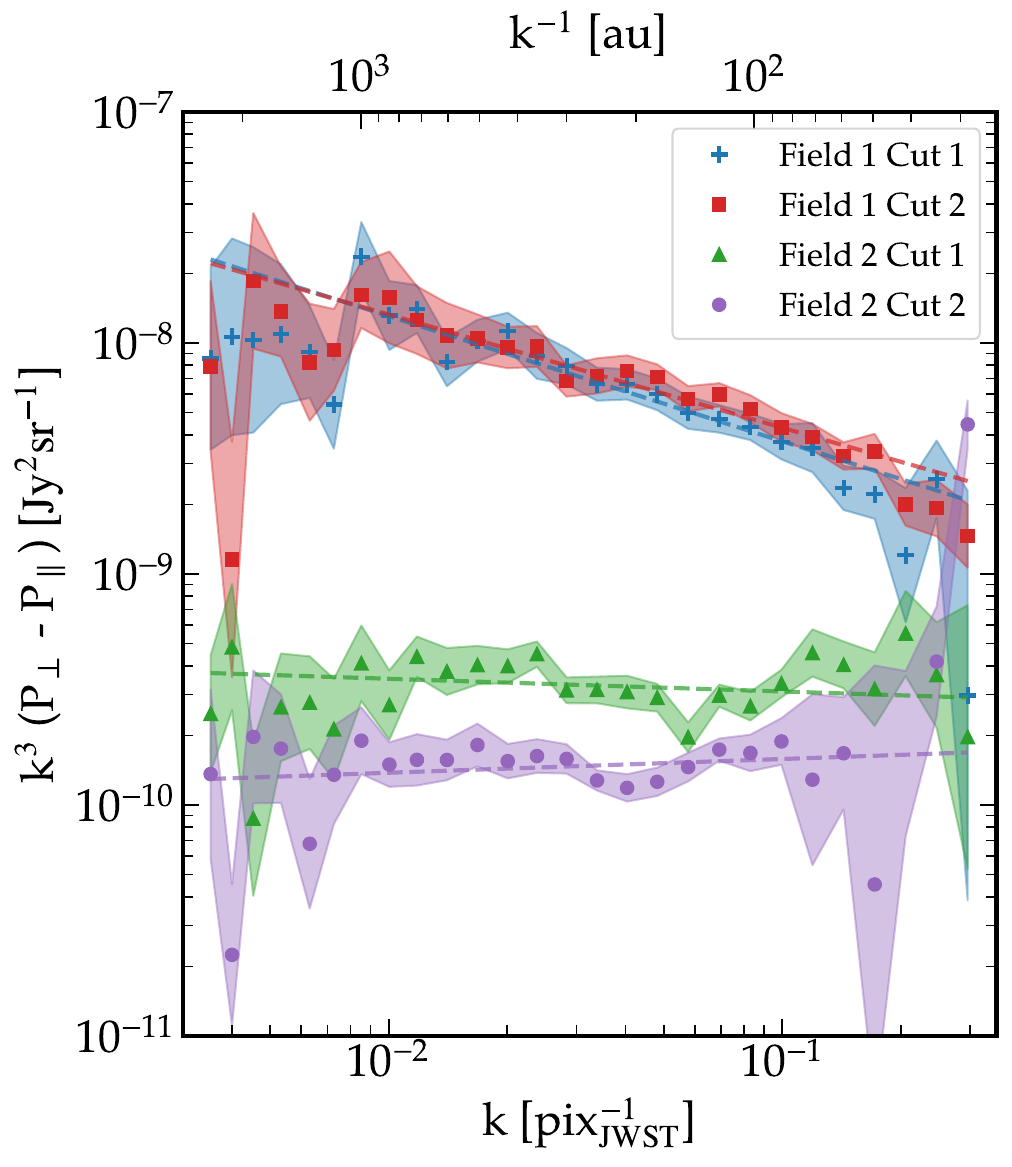}
\caption{Difference between the power spectra $P_\perp$ and  $P_\parallel $ for the four JWST images. The difference is scaled by $\mathrm{k}^3$ to reduce the dynamic range. The $1\,\sigma$  uncertainties are represented with shaded colors. The dashed lines represent least square fits to the spectra's differences. }
\label{fig:Delta_PS}
\end{figure}

\subsection{Power spectra from anisotropy}
\label{subsec:aniso}

In our second approach, we assume that the residual CIB is isotropic and obtain power spectra for each of the JWST images by taking the difference $P_\perp - P_\parallel$. These differences are shown in Fig.~\ref{fig:Delta_PS}.  To highlight the difference in slope between the Field~1 and 2 images, the spectra are scaled by $k^3$. The dashed lines represent power-law fits to the spectra:
\begin{equation}
P_{\perp} - P_{\parallel} =  A_a \left(\frac{k}{k_0}\right)^{\alpha_a},
\label{Eq:pwfit_dif}
\end{equation} 
The fit parameters $A_a$ and $\alpha_a$ for each of the images are listed in Table~\ref{tab:PS_aniso}. We also list values for the light and heavy masks to assess systematic errors. The differences with the values for the fiducial mask are minor.   
It is satisfying to obtain values for the power-law indices of the spectra, which are consistent with those obtained for the Field~1 cuts using the spatial difference (see Table~\ref{tab:PS1d}).

\subsection{Two-component fit}
\label{subsec:mcmc}

We present a third approach where we fit the $P_\perp$ and $P_\parallel$ spectra separately. In this fit, we assume that the CIB residual is isotropic and well approximated by a power law. The equations of the data model are as follows:
\begin{eqnarray}
\begin{aligned}
P_{\parallel} &= A_{0,\parallel} \, \, (k/k_0)^{\alpha_\parallel} + A_R \, (k/k_0)^{\alpha_R} \\
P_{\perp} &= A_{0,\perp} \, (k/k_0)^{\alpha_\perp} + A_R \, (k/k_0)^{\alpha_R},
\label{eq:pwfit_2D}
\end{aligned}
\end{eqnarray}
where the PAH power spectra are fitted with power laws that have distinct amplitudes and spectral indices.
\begin{figure}[ht]
\centering
\includegraphics[width = 235 pt]{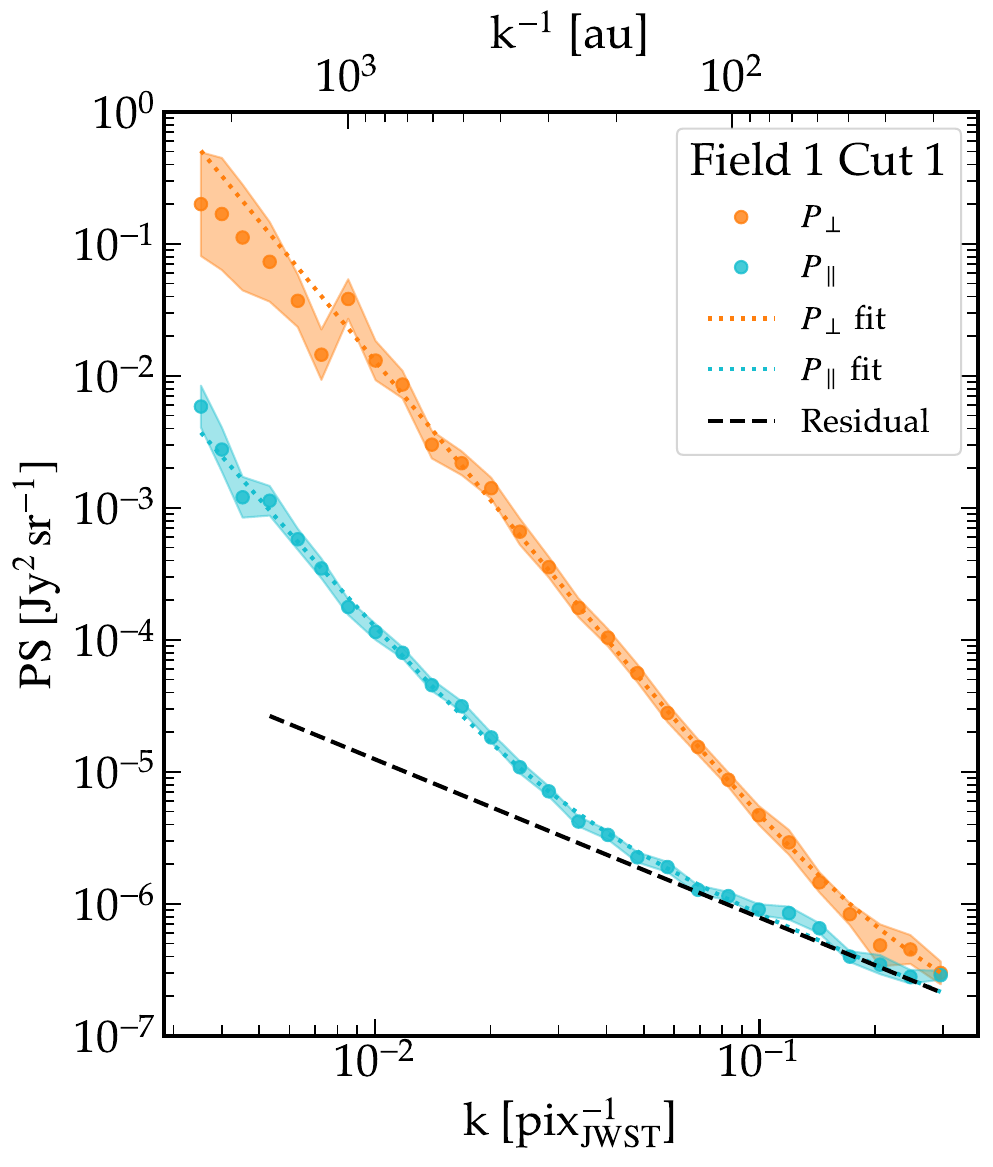}
\caption{Fit with two power laws of $P_\perp$ and $P_\parallel $ for the Field~1 image Cut~1.  The dotted orange and cyan lines represent the data fits of $P_\perp$ and  $P_\parallel $, respectively,  while the dashed black line represents the power-law fit of the CIB residual. The shaded areas around the data points indicate the $1\sigma $ error bars of the  spectra.}
\label{fig:spectra_multi}
\end{figure}

We performed a Monte Carlo Markov chain (MCMC) analysis to sample the joint posterior distribution of the model parameters. We utilized the EMCEE python package as described by \citet{Foreman13} in accordance with their recommendations. We used a Gaussian likelihood. The priors
force the three amplitudes $A_{0,\perp}$, $A_{0,\parallel}$, and $A_R$ to be positive, and the spectral indices $\alpha_\perp$ and $\alpha_\parallel$ to be negative. In addition, we set a Gaussian prior in the spectral index $\alpha_R = -1.2$ with a dispersion of 0.2 ($1\sigma$). This prior follows from the fit of $P_\parallel$ for Field~2 images illustrated in Fig.~\ref{fig:residual}. As this prior makes use of Field~2 data, we ran the MCMC analysis only for the two Field~1 images. 

\begin{figure}[ht]
\centering
\includegraphics[width = 235 pt]{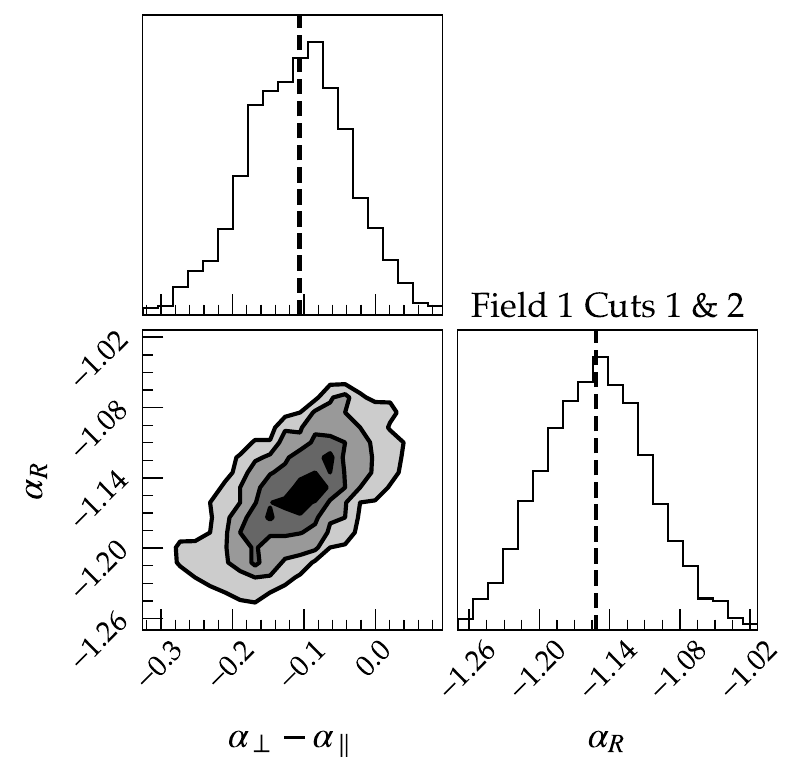}
\caption{Posterior  distribution of $\alpha_\perp - \alpha_\parallel$ and $\alpha_R$. The plot shows the joint distribution of $\alpha_\perp - \alpha_\parallel$ and $\alpha_R$ derived from the two power-laws fits of $P_\parallel$ and $P_\perp$ for Field~1 images Cuts 1 and 2 using the data model in Eq.~\ref{eq:pwfit_2D}. }
\label{fig:corner_plot}
\end{figure}

The model parameters with their error bars are listed in Table~\ref{tab:MCMC}. The table lists the median values of the posterior distribution and error bars computed from the 16 and 84 percentiles.
As shown in Fig.~\ref{fig:spectra_multi} for Field~1 Cut~1, the model provides a good fit for both $P_{\parallel}$ and $P_{\perp}$. The amplitude $A_R$ of the CIB residual is consistent with that measured for parallel spectra of Field~2, within  error bars (see Table~\ref{tab:PSfit_residual}). 
The values of $\alpha_\perp$ also match  the spectral indices  derived in Sects.~\ref{subsec:ON-OFF} and \ref{subsec:aniso}. 
It is satisfying to validate these values using a distinct approach in which the CIB residual is included in the data model with a power-law spectrum, rather than being subtracted by a difference. We note that the degree of anisotropy, measured by the ratios $P_{0,\perp}/P_{0,\parallel}$ in Table~\ref{tab:MCMC}, differs by a factor of 2 between Cuts~1 and 2 .  

Figure~\ref{fig:corner_plot} shows the joint posterior distributions of  $\alpha_\perp - \alpha_\parallel$ and $\alpha_R$, derived from the MCMC fits of Field~1 Cuts~1 and 2. 
The difference between the two spectral indices depends on $\alpha_R$ and, more generally, on the simplified model used to separate the PAH spectra from the CIB residual.  In Fig.~\ref{fig:corner_plot}, the median value of $\alpha_\perp - \alpha_\parallel$ is slightly negative, but we consider the deviation from zero to be not statistically significant. We conclude that the two spectral indices are identical within a measurement error bar of 0.2 ($3\sigma$). 

As \citet{Lithwick01}\footnote{See Sect.~3 in this publication} pointed out in the context of their interpretation of radio scintillation observations, the observational effects of the parallel direction are washed out due to variations in the local orientation of anisotropy along the line of sight and across the sky. In our observations, the PAH emission is so highly anisotropic that modes perpendicular to the local orientation of anisotropy dominate the Fourier spectra in all  projected orientations onto the plane of the sky. This statement entails three key facts. First, the ratios $P_{0,\perp}/P_{0,\parallel}$ in Table~\ref{tab:MCMC} are lower limits to the intrinsic anisotropy of the PAH emission. These limits depend on the dispersion of the local orientation of the anisotropy within the emitting medium. Second, the spectral index $\alpha_\parallel$  in Eq.~\ref{eq:pwfit_2D} does not represent the spectral index 
of the energy cascade in the direction parallel to the magnetic field. Third, the finding that $\alpha_\perp - \alpha_\parallel$ is consistent with 0 does not exclude the possibility that the spectral indices of the turbulent energy cascade for perpendicular and parallel modes differ. 

\begin{table*}[ht]
    \caption{MCMC fit of Field~1 images with two power laws.}
    \label{tab:MCMC}
    \begin{tabular}{lcccccc}
    \hline

    Image    & $P_{0,\perp} $    & $P_{0,\perp}/P_{0,\parallel} $    & $\alpha_\perp$ & $\alpha_\perp-\alpha_\parallel$ & $A_R$ & $\alpha_R$ \\
    & $\mathrm{Jy^2\,sr^{-1}}$  &  & & & $\mathrm{Jy^2\,sr^{-1}}$  & \\ 
    \hline 
    Cut~1   & $0.0115 \pm  0.0010$  & $92 \pm 9$ &  $-3.44 \pm 0.05$ &   & & \\ 
    Cut~2  & $0.0148 \pm 0.0012$  & $43 \pm 4$ & $ -3.54 \pm 0.04$ &   & &\\ 
    Cuts~1 \& 2   &  & & & $-0.10 \pm 0.07$ & $1.85 \pm 0.68 \,10^{-6}$   & $-1.15 \pm 0.05$ \\ 

    \hline
\end{tabular}
\\[1.0ex]
\tablefoot{Parameters of the joint fits of $P_\perp$ and $P_\parallel$ for Field 1 images Cuts~1 and 2, using data model in Eq.~\ref{eq:pwfit_2D} with $k_0 = 10^{-2}\,$pix$^{-1}$.}
\end{table*}

\section{Relation to interstellar turbulence}
\label{sec:turbulence}

In this section we present three complementary perspectives that relate the JWST Pleiades observations to interstellar turbulence. Together, they provide a framework that we hope will motivate theoretical studies aimed at interpreting the data.

\subsection{Dynamical interaction between the Pleiades stars and the surrounding ISM }
\label{subsec:interaction}

The Pleiades reflection nebulosity displays an impressive array of striations, which are locally nearly parallel. The origin of these striations remains a long-standing question. The idea that they could result from the dynamical interaction between the Pleiades stars with the surrounding ISM is not new. It was first discussed by \citet{Arny77}, and remains attractive. The dust and gas observations analyzed by \citet{Herbig01}  and \citet{Ritchey06} are direct evidence of a dynamical interaction. The change in dust polarization angles observed with the {\it Planck} data in Fig.~\ref{fig:Pleiades_Planck_polar} is further evidence of this.

Projection effects make it difficult to infer the distance at which the interaction with the stars occurs. Modeling of scattered light from dust indicates that the distance between matter and the brightest Pleiades stars ranges from a few tenths to 1 parsec. Specifically, it is a few tenths of a parsec near Merope \citep{Gibson03b}, a value that is consistent with our estimate of the radiation field intensity at the position of the Field~1 images (see Appendix~\ref{sec:context}). For Merope's luminosity, a value of G in the range 10–20 corresponds to a distance from the star between 0.3 and 0.5\,pc.     

\citet{Gordon84} discuss three ways in which the gas and stars could interact: radiation pressure on the dust, stellar winds, and thermal pressure of ionized gas. Each of these could possibly account for the lack of matter near the stars. The JWST observations show that the striations extend to scales three orders of magnitude smaller than the scale of the interaction.  Given that the Reynolds number of interstellar turbulence is high \citep{Elmegreen04}, what we observe with the JWST is most likely the signature of a turbulent energy cascade driven by interaction, rather than the interaction itself. Energy transfer could, for example, be mediated by radiation pressure, as investigated by \citet{Callies25}. 

\subsection{CNM structure on the scales of the Pleiades JWST observations}

The conventional framework for interpreting CNM observations \citep{Kritsuk18,Bellomi20,Marchal21} is based on numerical studies of the interplay between supersonic turbulence and thermal instability \citep{Seifried11,Kim13,Hennebelle14,Saury14}.
The JWST data open a novel perspective by tracing the structure of the CNM on spatial scales where the turbulence turnover timescale is shorter than the CNM cooling timescale. 

To support this statement, we computed the eddy turnover timescale for a power law scaling of the velocity dispersion: 
\begin{align}
\label{turnover_timescale}
 & v_\ell \, = \, v_L\, \left(\frac{\ell}{L}\right)^{\alpha_v} \\
 &   \tau_\mathrm{eddy} \, \equiv \,  \ell/v_\ell \sim \frac{\ell^{1-\alpha_v}\, L^{\alpha_v°°}}{v_L},
\end{align}
where $v_\ell$ is the dispersion of turbulent velocity on the scale $\ell$ and 
$L$ is the scale of energy injection.  
We used the size of the Pleiades nebula for $L$, and the dispersion of velocities among gas tracers reported by \citet{Ritchey06} as an observational estimate of $v_L$.

We assume that magnetic fields are dynamically important on the scale of the Pleiades nebula. We therefore used the canonical scaling of the velocity dispersion for MHD turbulence $\alpha_v = 1/4$ \citep{Iroshnikov64,Kraichnan65}.
For this index, the velocity dispersion is subsonic on the scales ($\ell \lesssim 10\,$mpc) statistically sampled by our JWST observations. The eddy turnover timescale is as follows:
\begin{equation}
 \tau_\mathrm{eddy} = 2\times 10^4 \, \left(\frac{\ell}{10\,\mathrm{mpc}}\right)^{3/4}\,\left(\frac{L}{3\,\mathrm{pc}}\right)^{1/4} \, \left(\frac{v_L}{2\,\mathrm{km s^{-1}}}\right)^{-1} \,yr. 
\end{equation}

We compared $\tau_\mathrm{eddy} $ with the gas cooling time, which we computed for a reference pressure in the local ISM $p/k = 4\times 10^3\,\mathrm{K\, cm^{-3}}$ using the analytical expression of the cooling function in \citet{Jennings21}. For $\ell < 10\,$mpc, $\tau_\mathrm{eddy}$ is at least two orders of magnitude smaller than the cooling timescale of the warm neutral medium. 
For $\ell \sim 10\,$mpc, $\tau_\mathrm{eddy}$ is comparable to the cooling time of the CNM, and shorter for lower values of $\ell$.
It is also relevant to estimate the Field length ($\lambda_F$). Thermal instability is suppressed by conduction on scales smaller than $\lambda_F$.
\begin{align}
\label{Field_length}
    \lambda_F \, & = \, 2\pi\,\sqrt{\kappa \, T / n^2 \Lambda} \\
        & = \, 30 \, (\kappa/10^5 \,\mathrm{erg s^{-1} cm^{-1}K^{-1}})^{0.5}\, \mathrm{mpc}. 
\end{align}
where $\kappa$ is the thermal conduction coefficient, $n$ the gas density,  $T$ the gas temperature, and $\Lambda$ the gas cooling rate \citep{Jennings21}. The numerical value is calculated for characteristic values of $n~(50 \,\mathrm{cm}^{-3})$ and $T~(80 \,\mathrm{K})$ in the CNM. We find that $\lambda_F$ is larger than 
the scales probed in our data analysis. 
These calculations suggest that the JWST observations of the Pleiades probe spatial scales on which turbulence in the CNM is  subsonic and is no longer coupled to the thermal instability. As such, the data offer a new angle for comparing observations and theoretical models.

\subsection{MHD subsonic turbulence}

In order to interpret our data in the context of turbulence, one would need to connect the observational findings of this work with theoretical models of MHD turbulence scaling laws and numerical simulations \citep[see][for a recent review]{Schekochihin22}. This is obviously a difficult task and it is beyond the scope of our observational analysis to draw conclusions about specific models.  Our aim is to guide future theoretical investigations. With this in mind, we present what we consider to be the four key observational results of our study. 
\begin{itemize}

\item 
The mean orientation of the map striations coincides with that of the mean magnetic field  projected onto the plane of the sky, as determined by the {\it Planck} dust polarization observations. 

\item 
The anisotropy power ratios relative to this mean orientation, 90 and 40 for Fields~1 and 2 (Table~\ref{tab:MCMC}), are lower bounds to the local anisotropy, due to the dispersion in the orientation of the striations (Sect.~\ref{subsec:aniso_model}). This dispersion -- as determined by the  widths of the angular profiles of the 2D power spectra $P_{2D}$ -- is significant (see Table~\ref{tab:angles}). 

\item
The anisotropy of $P_{2D}$ is  observed to be scale-independent within a small uncertainty set by the separation between the PAH emission and the CIB residual. We stress that for a large power anisotropy, we expect perpendicular modes to be dominant in all  orientations projected onto the plane of the sky (Sect.~\ref{subsec:mcmc})

\item
The spectral indices of the PAH power spectra measured for Field~1 match the reference value -3.5 for MHD turbulence. However, we note that the values for Field~2 differ significantly from this reference. 

\end{itemize}

We discuss two possible interpretations of the origin of the power anisotropy. This  choice is illustrative and not exclusive of other models. 
(1) Striations are formed by fast magnetosonic waves traveling perpendicularly to the mean magnetic field.
(2) They are fingerprints of the anisotropy of the turbulent energy cascade of subsonic MHD turbulence. 
In the first case, the anisotropy is global with respect to an average guide field, while the anisotropy of the turbulence energy cascade occurs with respect to the local orientation of the magnetic field. 

The first interpretation was introduced and quantified using numerical simulation in the context of molecular cloud observations by \citet{Tritsis16} and \citet{Tritsis18}. It was also investigated by \citet{Beattie20} using numerical simulations. More research involving dedicated simulations is required to verify whether this interpretation can reproduce the main observational results. 
The second is based on the \citet{Goldreich95} model of subsonic MHD turbulence. In this framework, compressible modes are sheared and passively mixed by Alfv\'en waves \citep{Lithwick01}. This interpretation is attractive because it links the JWST observations to the anisotropic nature of the MHD turbulence energy cascade \citep{Boldyrev06,Schekochihin22}.

\section{Conclusion}
\label{sec:conclusion}

We have presented and analyzed NIRCam JWST observations of the Pleiades nebula in the F335M filter centered on the PAH emission feature at $3.3\, \mu$m. 
These JWST observations provide an opportunity to analyze the structure of the CNM in exceptional detail, reaching spatial scales where existing observations are too limited to enable a statistical analysis.
The main results of our work are summarized as follows. 

We built maps of diffuse PAH emission, separating it from sources, stars, and galaxies, as well as from the extended spikes of bright stars. The stars were subtracted using an empirical template of the JWST PSF. Galaxies were identified and masked using multi-resolution image filtering to interpolate the structure of diffuse emission. 

We have presented a power spectrum analysis of the diffuse emission in four images, two of which are close to the Merope star and two of which are more distant. The spectra exhibit flattening at high wavenumbers, which we identify as a CIB residual. The spectral index of this component, -1.2, is similar to that measured for galaxy clustering in CIB observations at near- and far-infrared wavelengths. 
    
The amplitude of PAH power spectra are highly anisotropic on all scales. We characterized them in three independent ways: (1) We subtracted the CIB residual by taking the difference of the spectra computed for the JWST images near and far from Merope. (2) We differentiated anisotropic PAH emission from the isotropic CIB residual by modeling the 2D angular distribution of Fourier power. (3) We fit the power spectra in the directions parallel and perpendicular to the anisotropy orientation as the sum of two power laws, representing the contributions from PAHs and the CIB. The three methods yield consistent results. 

The PAH power spectra are measured to a spatial resolution of 0.2 \,mpc (40 \,au). They are well fitted with single power laws, with no sign of a break that could be interpreted as a signature of a characteristic scale associated with turbulence dissipation. The power-law indices are -3.5 in the two fields near Merope and -3 in the more distant areas. The magnetic field orientation inferred from  dust polarization data from {\it Planck } is aligned with the PAH anisotropy. The power anisotropy does not change with scale. 

These findings are discussed in the context of interstellar turbulence, which may be driven by the dynamical interaction between the Pleiades stars and the surrounding ISM. The observations offer a new perspective for comparing observations and theoretical models because they reveal physical scales at which CNM turbulence is decoupled from the thermal instability. 
Progress in interpreting the data within this framework requires follow-up theoretical work, including numerical simulations to link the PAH power spectrum to the 3D anisotropic structure of the magnetized ISM.

\begin{acknowledgements}
GV, AAC and NDR thank the STScI for DDRF Summer funding to support their research. We thank the referee for a detailed report, which helped us to clarify the discussion of the results of the data analysis in relation to interstellar turbulence. This work is based on observations made with the NASA/ESA/CSA James Webb Space Telescope which can be found in MAST: \href{https://doi.org/10.17909/26am-1s41}{10.17909/26am-1s4}. The data were obtained from the Mikulski Archive for Space Telescopes at the Space Telescope Science Institute, which is operated by the Association of Universities for Research in Astronomy, Inc., under NASA contract NAS 5-03127 for JWST. These observations are associated with program 2143.
\end{acknowledgements}

\bibliographystyle{aa} 
\bibliography{main} 

\appendix
\section{Obtaining images of the diffuse emission  } 
\label{App:data_cleaning}

\begin{figure}[ht]
            \centering
            \includegraphics[width = 250 pt]{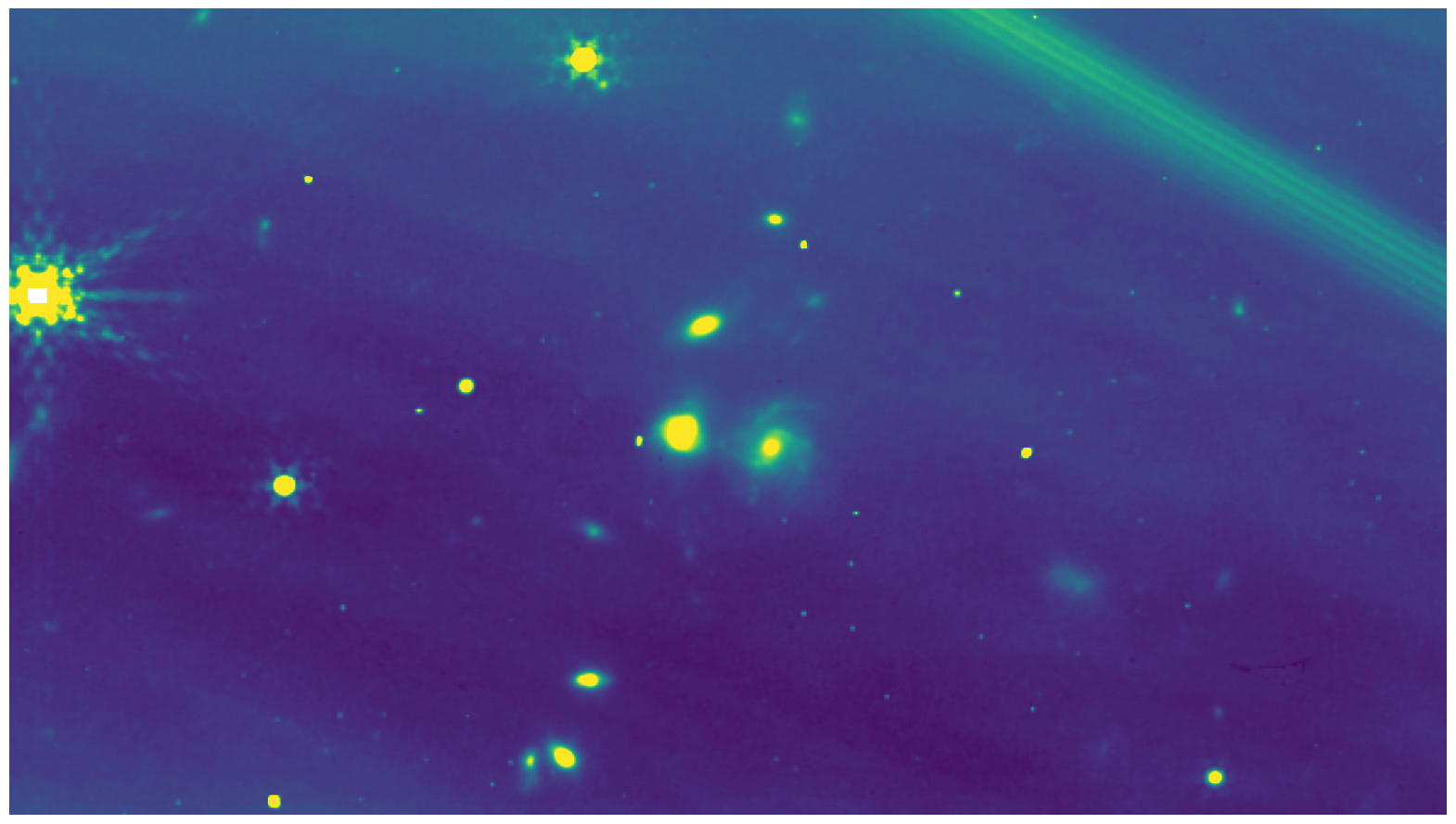}
            \caption{A region, about $40\arcsec \times 20\arcsec$, of the Field~1 mosaic illustrating the diversity in emission sources in our observations: diffraction spikes, saturated stars, unsaturated stars, glitches, and galaxies of various shapes and sizes.}   
            \label{fig:4galaxies}
        \end{figure}

        This appendix details and illustrates the methods used to create maps of diffuse emission out of the JWST mosaics. Figure~\ref{fig:4galaxies} presents a zoomed-in view of the data to illustrate the diversity of emission sources we needed to subtract or mask to obtain maps that can be used to statistically analyze the diffuse emission. 
        
    \subsection{Removing diffraction spikes from off-field point sources}
    \label{itm: removemerope}

        The bright star Merope in the vicinity of our field of view produces three diffraction spikes that are bright features across Field~1 (see the bottom-left panel of Fig.~\ref{fig:pleiades}). 
        The central one to the east of Merope aligns with the NIRCam detector arrays, and the other two are at $60^\circ$. 
        The Merope spike with the lowest declination extends to Field~2. We modeled and removed the three spikes from Merope in Field~1, the spike from Merope that extends to Field~2, and two fainter spikes from PQ Tau in Field~1. The first three spikes are clearly visible on the bottom right panel in Fig.~\ref{fig:pleiades}, while the fourth is barely noticeable on the bottom left panel. The last two are too faint to be visible in the figure. We illustrate our modeling with the eastern spike. To model the oblique spikes, we rotated the JWST mosaic in array coordinates by plus or minus 60 degrees to apply the same procedure.
        
\begin{figure}[ht]
        \centering
        \includegraphics[width = 250 pt]{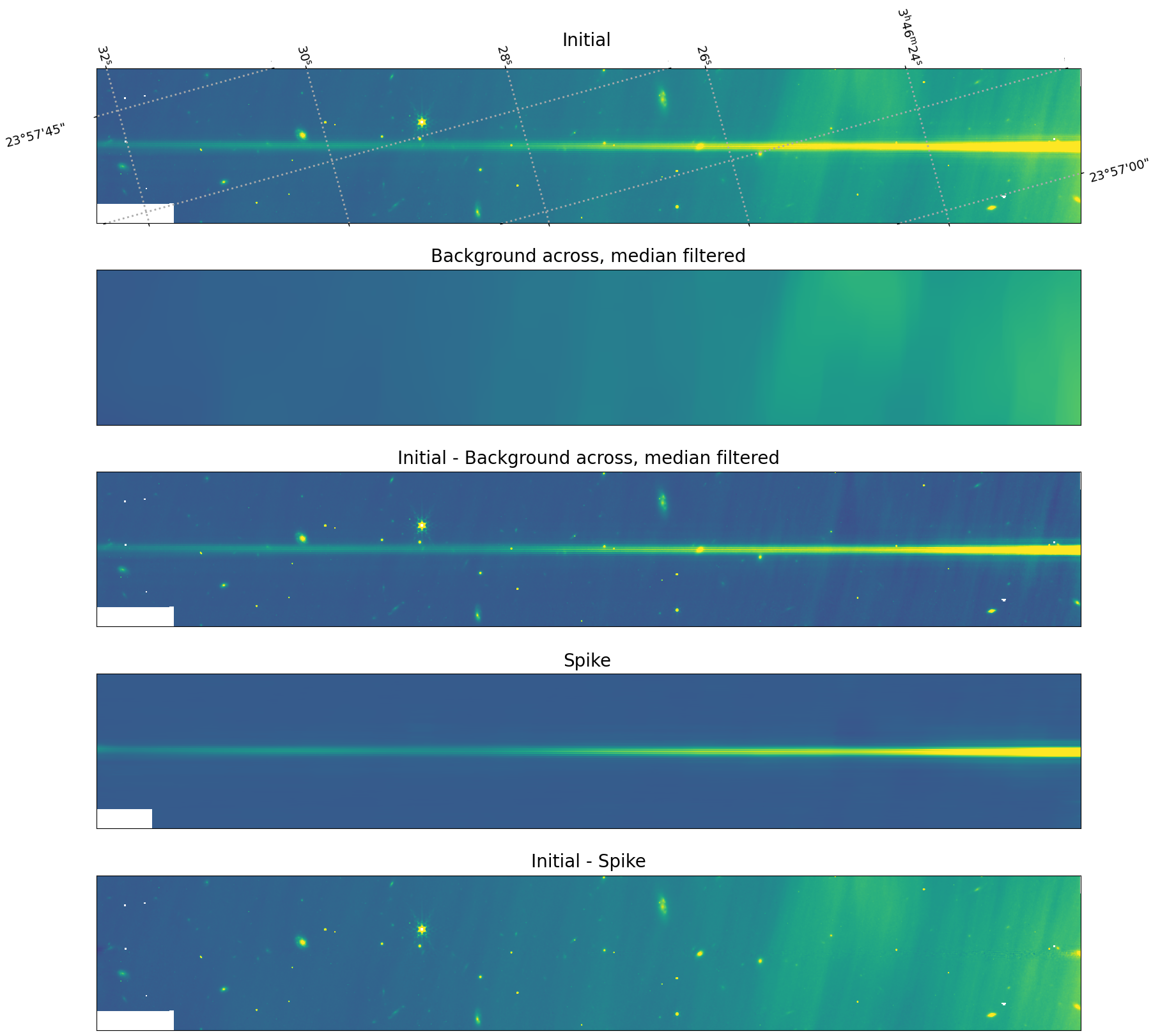}
        \caption{Modeling and subtraction of the spike to the east of Merope. The five panels show the same image cut out at different steps of the data cleaning (from the top): the original map; the estimation of the background with median spatial filtering; the original map with the background subtracted; the  spike model; and the initial map with the spike model subtracted. 
        A nonlinear color scaling is used to enhance the low-surface-brightness features in the field.}
        \label{fig:arm_removal}
    \end{figure}

        We cut a rectangular section of the observations surrounding the diffraction spike, which is 75 to 175 pixels wide on each side (see the first panel in Fig.~\ref{fig:arm_removal}). The width chosen depends on the intensity of the spike. In the following, columns refer to 1D cuts within the cutout image, across the spike, and rows to 1D cuts that are parallel to the spike. The steps involved in modeling and subtracting the spikes are illustrated in Fig.~\ref{fig:arm_removal}.

        The first step is to estimate the background emission around the spike. To achieve this, we masked the inner 50 to 150 pixels where the spike dominates the emission. We also masked any value above $2\,\sigma$ from the median in each column (using \href{https://docs.astropy.org/en/latest/api/astropy.stats.sigma_clip.html}{\texttt{sigma\_clip}} from \texttt{Astropy}). The data along each column are fitted with a third-order polynomial function, which we used to interpolate and fill in the diffuse emission at the position of the masked pixels. We applied a median filter to the polynomial interpolation, with a width along the spike orientation of 100 pixels, to filter out irregularities due to compact sources. The resulting map is shown in the second panel of Fig.~\ref{fig:arm_removal}.
        We subtracted this background from the initial cutout image, which yielded a map of the spike. This map still contains point and compact sources, as well as some residual diffuse emission (see the third panel in Fig.~\ref{fig:arm_removal}).

\begin{figure}[ht]
            \centering
            \includegraphics[width = 250 pt]{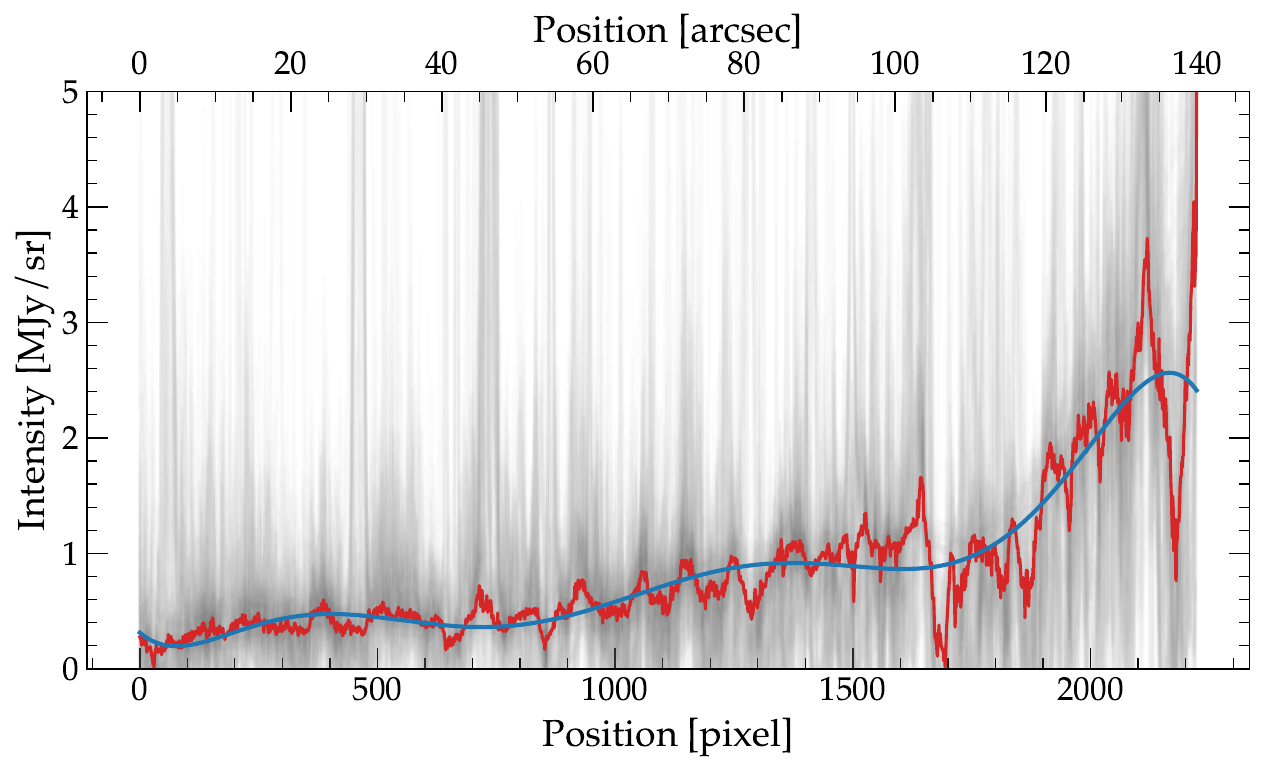}
            \caption{Intensity gradient along the  spike to the east of Merope. In gray are all the individual longitudinal profiles within a 150-pixel-wide box around the spike, after subtraction of the diffuse background estimate and normalization across the spike (see text for details). The $2\,\sigma$-clipped median profile is plotted in red, and the fit to the median using a seventh-order polynomial function is plotted in blue.}
            \label{fig:longitudinal_fit}
        \end{figure}

        The second step involves modeling the spike. We estimated the gradient along the diffraction spike assuming that it can be modeled by one polynomial function that applies to each row of pixels. To obtain this polynomial function, we calculated the $2\,\sigma$-clipped median of each column across the width of the spike: 150 pixels for the eastern spike of Merope, 100 pixels for the other spikes. This gave us the typical longitudinal gradient profile of the spike, which a fit with a seventh-order polynomial function (see Fig.~\ref{fig:longitudinal_fit}).

\begin{figure}[ht]
            \centering
            \includegraphics[width = 250 pt]{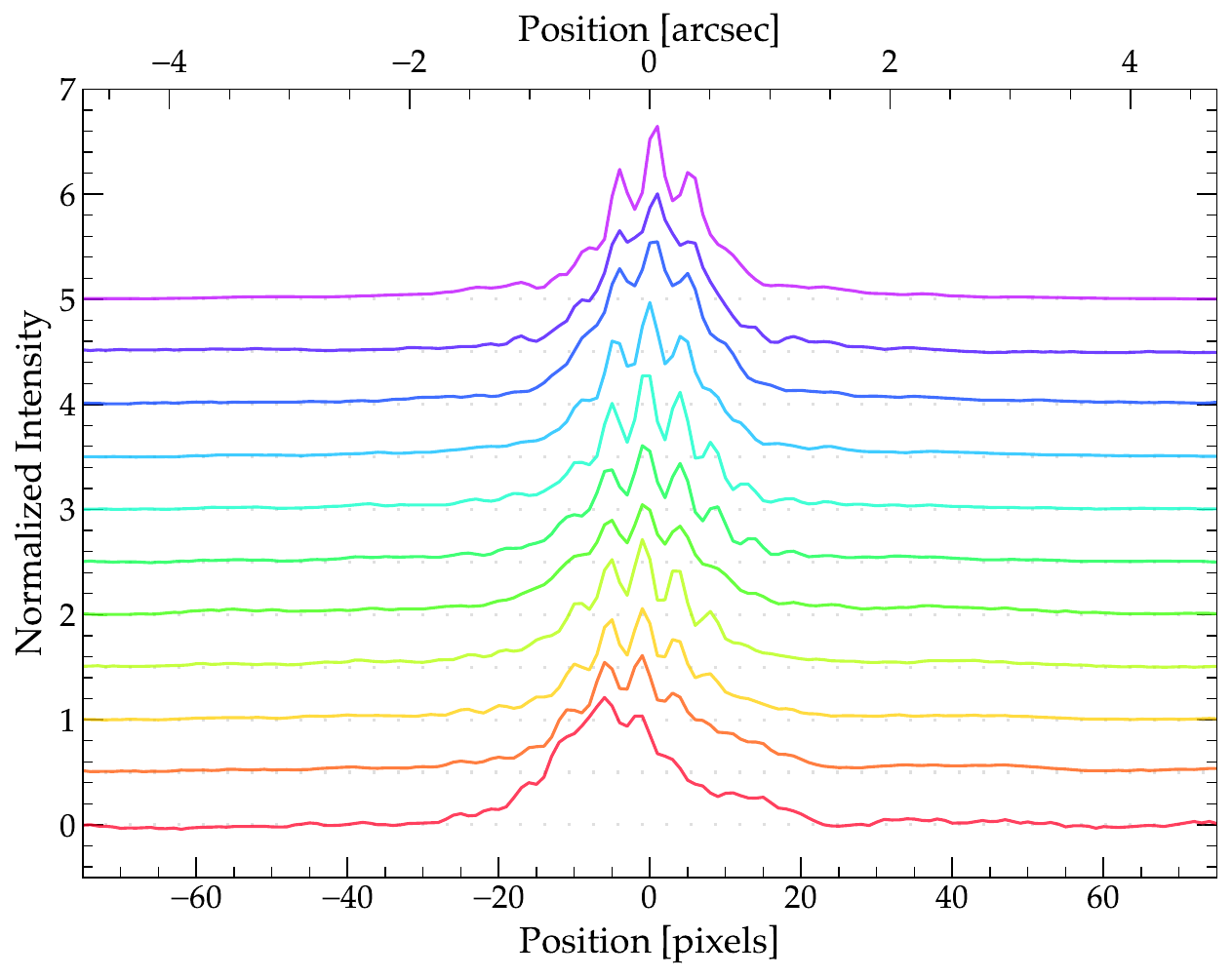}
            \caption{Emission profiles across the spike at multiple positions along the spike (every 200 pixels, or about 12.6\arcsec). Each profile is offset by 0.5 with respect to the previous one. The top profile is that closest to Merope.}
            \label{fig:horizonal_pattern}
        \end{figure}

        Finally, we normalized the first approximation of the spike we obtained earlier by dividing it, for each column across the spike, by the polynomial values of the longitudinal gradient. For each column, we then selected the nearest 100 columns to compute the $2\,\sigma$-clipped median of this selection for each row. This gave us the profiles across the spike, assuming it does not vary on scales much shorter than 100 pixels or 6.3\arcsec (see Fig.~\ref{fig:horizonal_pattern}). 

\begin{figure}[ht]
            \centering
            \includegraphics[width = 250 pt]{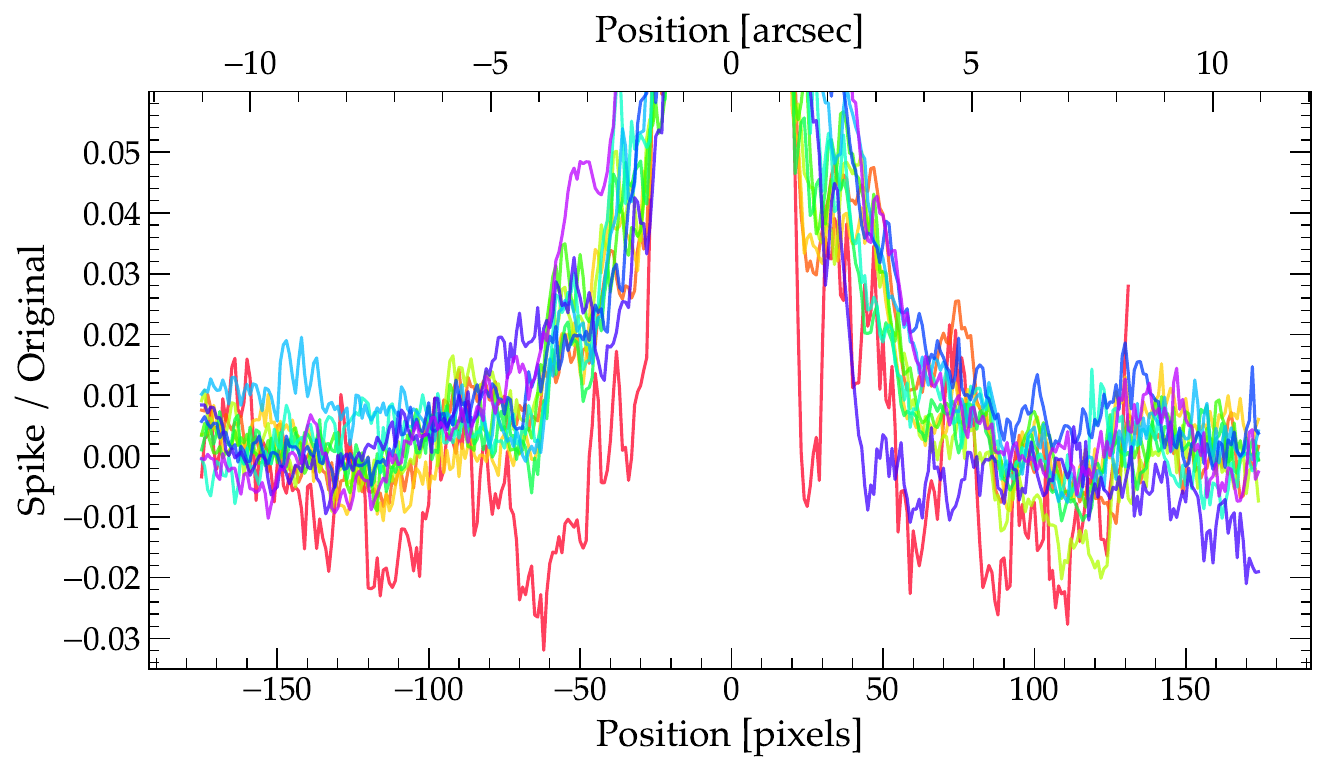}
            \caption{Profiles across the final spike at multiple positions along the spike (every 200 pixels, or about 12.6\arcsec), divided by the original image. The color code is the same as in Fig.~\ref{fig:horizonal_pattern}.}
            \label{fig:quality_spike}
        \end{figure}

        We note that the pattern varies smoothly along the spike. In particular, we see the peak of the profile shifts by about 2 pixels over 2000 pixels. A larger shift seems to occur at the faintest end of the spike, far from Merope, where the relative intensity of the spike is the lowest. However, even there, the shift is not greater than 7 pixels compared to the other end of the spike.
        Next, we rescaled the normalized profiles with the polynomial gradient to obtain the spike model shown in the fourth panel of Fig.~\ref{fig:arm_removal}.
        Finally, we subtracted the model from the initial map (see the fifth panel in Fig.~\ref{fig:arm_removal}). 
        To assess the accuracy with which the spike is removed, we compared the residual background in the spike image, away from the central area, relative to the initial map. We find that, in the case of Merope's straight spike, the residual is typically lower than 1\% at more than 75 pixels (4.7\,\arcsec) from the center of the spike (see Fig.~\ref{fig:quality_spike}). 

\begin{figure}[ht]
            \centering
            \includegraphics[width = 200 pt]{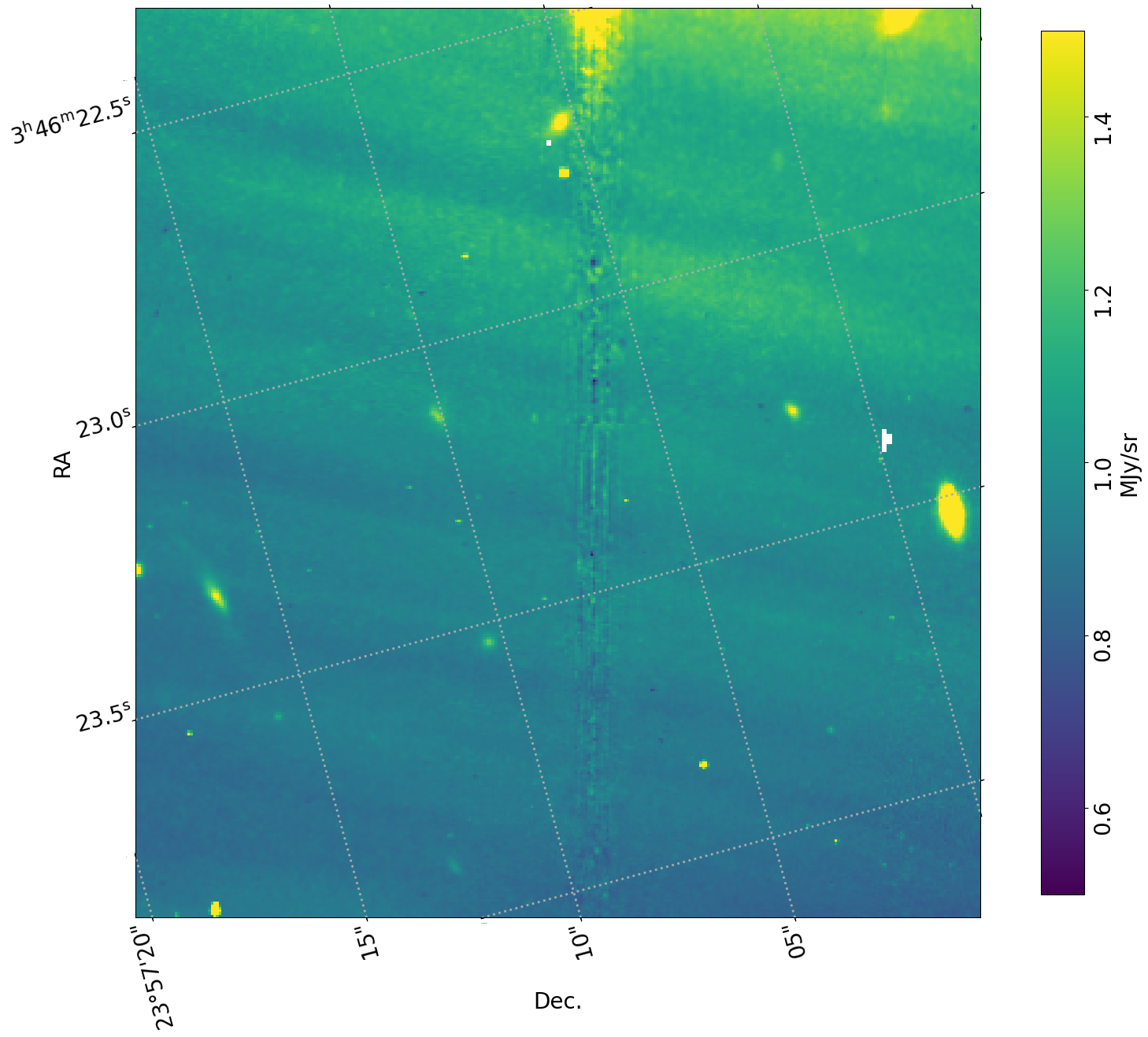}
            \caption{Residual of the spike to the east of Merope. The color scale has been stretched relative to other figures to enhance the faint residual.}   
            \label{fig:arm_issue}
        \end{figure}

\begin{figure}[ht]
            \centering
            \includegraphics[width = 250 pt]{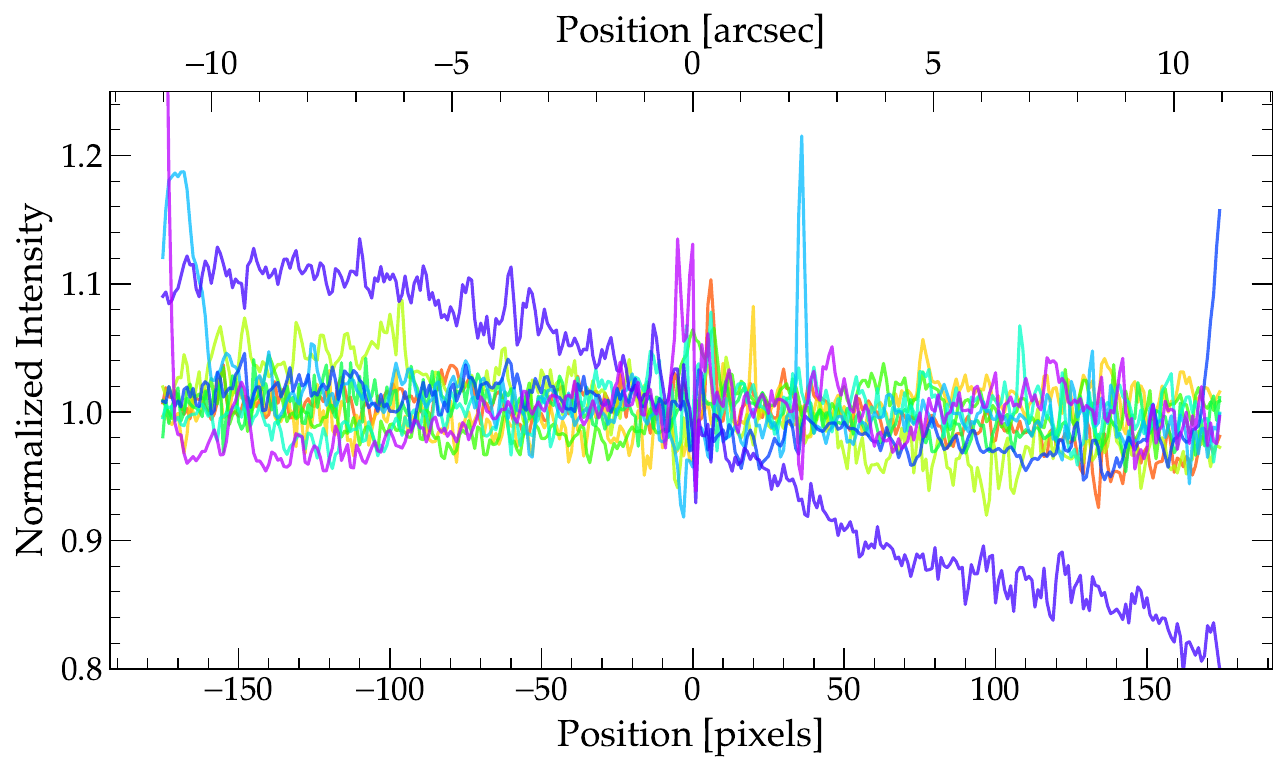}
            \caption{Profiles across the location of the spike at multiple positions along the spike (every 200 pixels, or about 12.6\arcsec), obtained on the final image, normalized to equalize their median values. The color code is the same as in Fig.~\ref{fig:horizonal_pattern}.}
            \label{fig:quality_background}
        \end{figure}

        A limitation in our method is visible in Fig.~\ref{fig:arm_issue}. Closer to Merope, the pattern across the spike varies on scales smaller than 100 pixels (or 6.1\arcsec), leaving a residual trace near the edge of our observations. However, the alternative (i.e., using a smaller window than 100 pixels) would not properly filter out the sources ``hidden'' by the spikes. We compared these residuals with local dust emission by normalizing the profiles at the site of the spike to their median values (see Fig. \ref{fig:quality_background}). In most cases, the residual of the spike is barely noticeable above the intrinsic variations of the diffuse emission, at a level of 5\% of the median value of the local background. However, near Merope, the residual reaches more than 10\% of that level but remains mostly random.

        At this point, we thus have four mosaics, two for each field, with all large diffraction spikes removed. For each pair of mosaics, we removed the glitches by comparing the two mosaics. Each pixel whose value varies by more than 0.5~MJy/sr was replaced by the lower value in the pair.

\subsection{Multi-scale background}

        Once the large spikes have been removed from the data, we modeled the diffuse emission.
        We used mosaics of Fields~1 and 2 from which the spikes have been removed. We computed their global median by clipping pixels above 2$\sigma$ and subtracted the result from the images. 
        
\begin{figure}[ht]
            \centering
            \includegraphics[width = 250 pt]{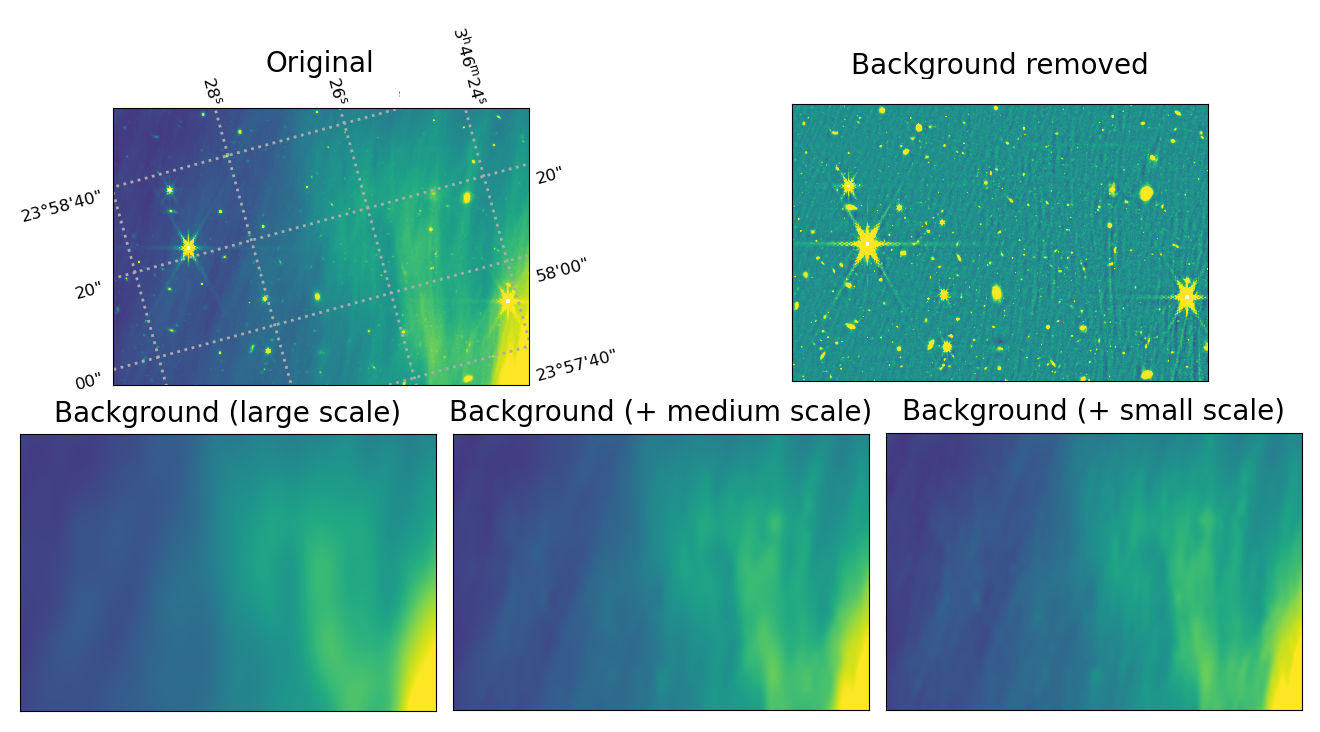}
            \caption{ Images illustrating the multi-scale composition of the background map on a sub-image within Field~1. From the top left and going anticlockwise, the five panels show: the sub-image with the spikes removed; the background map at 100-pixel scale including the global mean; the previous map with the addition of the 40-pixel scale background; the previous map with the addition of the 25-pixel scale background; and the residual image obtained by subtracting the multi-scale background from the original image. The color scale ranges from 0.35 to 0.95 MJy/sr except for the top-right panel, where it ranges from -0.05 to 0.05 MJy/sr. A nonlinear color scaling  is used to enhance the low-surface-brightness features in the field.}   
            \label{fig:bkg_multi}
        \end{figure}

        Next, we subtracted the different scales of the diffuse emission by repeating three times the following steps: (1) we computed the standard deviation of the 2$\sigma$-clipped background subtracted image, (2) we masked the bright pixels whose values are more than 15$\sigma$ above the median of the entire image, (3) we performed a 2D median-filtering with a 4$\sigma$ clipping without the masked pixels, and (4) we subtracted the median-filtered image.
        We used the \href{https://photutils.readthedocs.io/en/stable/api/photutils.background.Background2D.html}{\texttt{Background2D}} class from \texttt{photutils} to perform these steps \citep{photutils}.
        We applied the 2D median filter using rectangular boxes elongated along the direction of the detector arrays, which is close to the orientation of the striations.
        We used boxes with mean sizes of 100, 40, and 25 pixels. The ratio between the long and short sizes is 2 at all three scales.  Figure \ref{fig:bkg_multi} illustrates the composition of the background map as smaller scales are included, as well as the residual map obtained after background subtraction for a sub-image of Field~1. There is structure in the diffuse emission at scales smaller than 25 pixels (i.e., 1.5 arcsec). This contributes noticeably, albeit minimally, to the residuals, even in Field 1, where the diffuse emission is strongest.

\subsection{Removing point sources within the field}

        After subtracting the multi-scale background, we used the residual images to identify point sources and model the PSF.
        We  identified point sources  by recursively selecting the brightest pixel in the four combined mosaics. For each such pixel, we looked for all other neighbor bright pixels and labeled the cluster as a source before looking for the next brightest pixel in the observations. We extracted in this way the brightest 400 sources from our observations. We then visually inspected them all to separate point sources from extended sources, and ended up with 175 point sources.
        We did not use publicly available source extraction algorithms as they tend to identify multiple sources within each JWST PSF, as those are highly structured for the brightest sources. Our method is more robust and still functional  given the small number of sources we wanted to extract.
        
         \begin{figure*}[ht]
            \centering
            \includegraphics[width=\linewidth]{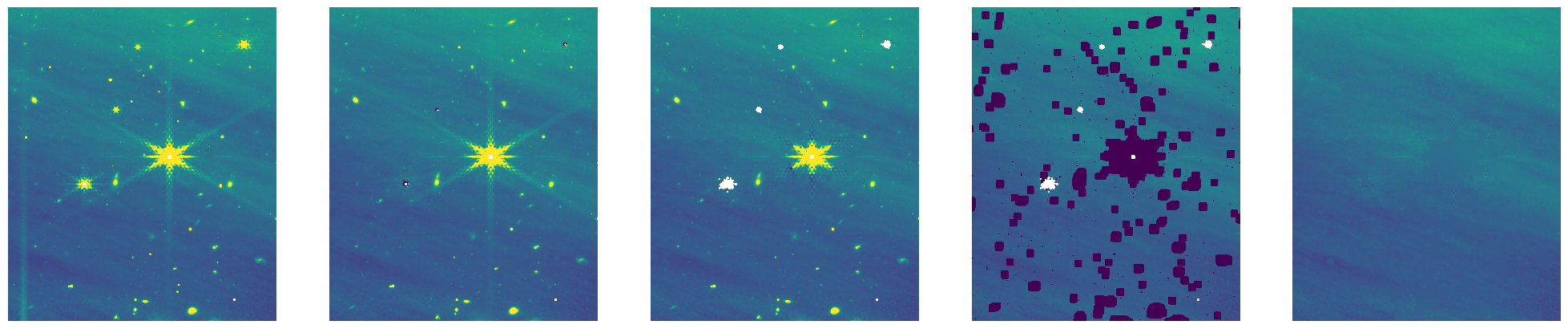}
            \caption{ Images illustrating  steps in the data cleaning on a sub-image of Field~1. From left to right: the initial image, which includes the spike of PQ~Tau close to the image edge to the left;  the  image after the spike from PQ~Tau and the in-field stars have been subtracted via the ePSF method (four stars mostly in the top half of the image); the subsequent image obtained by masking  the cores of the point sources; the same region after we mask all pixels that correspond to galaxies, faint stars, and hot pixels; and the diffuse emission map.}
            \label{fig:stars_removed}
        \end{figure*}
         
        To build an empirical PSF (ePSF), we followed the method developed by \citet{Anderson2000} and \citet{Anderson2016}, which is implemented as the \texttt{EPSFBuilder} class in \texttt{photutils}, adapting it to our specific needs. In particular, we wanted to clean the spikes of the point sources more thoroughly than their cores because the spikes contaminate the diffuse emission more than the cores.

        To combine the images of the selected point sources, we first centered each image to sub-pixel accuracy using the theoretical PSF of \texttt{WebbPSF} \citep{Perrin2014} on a 301 by 301 pixel ($\sim 20\arcsec$ by $20\arcsec$). Each PSF is fitted in a squared annulus region, instead of its core, and we find that using a region of 45 by 45 pixels, with the 9 by 9 innermost pixels masked, gives the best results. The difference between the position of the brightest pixel and the computed centers of the PSFs is typically less than 0.4 pixel.
    
        The centered PSFs are then stacked in an iterative way by our modified \texttt{EPSFBuilder} to generate the ePSF for our observations. Here, we used the same square annulus as before to align the individual PSFs with each other. The method also returns the individual best fit for each source in our collection, in their 301 by 301 original pixel grid, which we subtracted from the observations, except for the six brightest point sources that have diffraction spikes extending beyond 20\arcsec\ (see Fig.~\ref{fig:stars_removed}). There are four point sources in Field~1 and two in Field~2, for which the core is highly saturated. If we had removed their ePSF best fit, this would have led to significant residuals across the entire 301 by 301 pixel region. For these sources, we instead waited until we added back the multi-scale background, after which we removed their diffraction spikes in the same way as for the stars off field (see Appendix~\ref{itm: removemerope}).

    \subsection{Masking the remaining sources}
        At this stage, we had  removed all diffraction spikes from sources within and without our field of observations. We were  left with masking the saturated core of bright point sources and  compact sources.

        To build the mask of the remaining sources, we first identified pixels that correspond to significant residuals after subtracting the ePSF fit. We computed the 2$\sigma$-clipped standard deviation for each 301 by 301 pixel image of a point source, keeping only values above a floor of 0.1~MJy/sr. Any pixel in the image whose value in the residual of the ePSF fit is greater than 10 times this standard deviation is flagged. Any pixel whose value in the ePSF fit is greater than 15 times this standard deviation is also flagged. These pixels represent fewer than 0.5\% of all pixels in the four sub-mosaics and are the first included in a mask.

        Next, we masked additional bright pixels. We started by creating a multi-scale background in an identical manner to what was done in the previous subsection, but now starting from a cleaner map. We then selected all pixels where the residual emission after subtraction of the multi-scale background is higher than a brightness threshold (0.035~MJy/sr  for the fiducial mask). We then identified small and large clusters of masked pixels: those with fewer than or more than nine connected masked pixels. We verified that the small clusters typically correspond to faint stars or isolated glitches. Large clusters correspond either to the core of galaxies, whose shapes and sizes are very diverse (see Fig. \ref{fig:4galaxies}), or to the few bright point sources for which the ePSF method was not applied. Since the brightness threshold leaves the edges of bright sources unmasked, we extended the large clusters by a buffer zone of 15 additional pixels. To do this, we used the \texttt{skimage.morphology} package \citep{vanderWalt2014}. Finally, we  added back the small clusters of pixels to the mask.

        Starting from mosaics containing large diffraction spikes, point sources, and extended sources, we removed the spikes step by step  from both the off-field and in-field sources and masked both the core of the point sources and the extended sources. This left us with mosaics containing only diffuse emission, which were used to perform the Fourier analysis presented in Sect.~\ref{sec:Fourier}.

\section{Magnetized interstellar matter near the Pleiades stars} 
\label{App:interstellar}

\begin{table*}[ht]
\caption{\textit{Planck} and WISE measurements. }
\label{tab:Planck_dust}
\begin{tabular}{lcccccc}
\hline 
    Field   & $T_d$ & $\tau_{353}$  & $I$ & $Q$ & $U$ &$I_{12\mu \mathrm{m}}$\\
    
    & K  &   &MJy\,sr$^{-1}$ &MJy\,sr$^{-1}$ &MJy\,sr$^{-1}$ & MJy\,sr$^{-1}$ \\ 
    & (a) & (b) & (c ) & (c) & (c) & (d)\\
    \hline 
    {\bf Field~1} & & &    & & & \\
    Cut~1  &  $27.6 \pm 1.2$ &  $13.9 \pm 0.8 \, 10^{-6}$  & $0.63\pm 0.14$   & $0.076\pm 0.025$  & $-0.041\pm 0.025$ & $7.3  \pm 0.1$ \\ 
    Cut~2  &  $27.5 \pm 1.2$  & $14.5 \pm 0.8 \,10^{-6}$   &  $0.67\pm 0.14$  & $0.069\pm 0.025$  & $-0.056\pm 0.025$ & $6.1 \pm 0.1$ \\ 
    \hline
    {\bf Field~2} & & & & & & \\
    Cut~1  &  $26.7 \pm 0.2$ & $14.4 \pm 0.2 \,10^{-6}$  &  $0.66\pm 0.14$  & $0.038\pm 0.025$ & $-0.105\pm 0.025$ & $3.3 \pm 0.1$\\ 
    Cut~2  & $25.5 \pm 0.2$  & $13.8 \pm 0.2 \,10^{-6}$  &  $1.07\pm 0.14$  & $0.035\pm 0.025$ & $-0.122\pm 0.025$ & $2.4 \pm 0.1$ \\ 
    \hline
    {\bf Background$^e$} & $18.6 \pm 0.5$ & $12.3 \pm 3 \,10^{-6}$ & $0.58 \pm 0.14$ & $0.055\pm 0.025$ & $0.016\pm 0.025$ & $0.34 \pm 0.10$ \\
    \hline
    \end{tabular}

\tablefoot{(a) Dust temperature from {\it Planck} data at 5' resolution. (b) Dust opacity at $353\,$GHz from {\it Planck} data at 5' resolution. (c) Stokes $I$, $Q$ and $U$ at 353\,GHz measured at the center of the JWST images on {\it Planck} maps smoothed to a $7^\prime$ beam. The values for Fields~1 and 2 are background subtracted. (d) Brightness at $12\,\mu$m measured at the center of the JWST images on the WISE all-sky map \citep{Meisner14}. The values for Fields~1 and 2 are background subtracted. (e) Median values over the background area. The error bars are the standard deviation of pixel values. Uncertainties on the median values are one order of magnitude smaller.}  
\end{table*}

This appendix provides a broader context for the JWST observations. {\it Planck} and WISE data are used to characterize the magnetized interstellar matter associated with the Pleiades nebula. 

We used the dust temperature and opacity maps derived from fitting the dust spectral energy distribution at far-infrared wavelengths to a modified black body \citep{PIPXLVIII}, and the dust polarization maps at 353\,GHz \citep{Planck_LegacyXII}. We also used the full sky image produced by \citet{Meisner14} from the WISE data at $12\,\mu $m\footnote{Image downloaded from \url{https://faun.rc.fas.harvard.edu/ameisner/wssa/healpix.html}}. The angular resolution of these maps is 5'. However, Stokes $Q$ and $U$ are smoothed to 7' in order to improve the signal-to-noise ratio of the dust polarization. 

The {\it Planck} and WISE maps offer complementary insights into the ISM. The dust temperature traces the localized heating of dust grains by the Pleiades stars, while the dust opacity indicates the gas column density. The conversion factor from $\tau_{353}$ to $N_H$ for translucent molecular gas is $10^{26}\,\mathrm{H\,cm}^{-2}$ \citep{Planck14_XI}. The dust polarization angles are perpendicular to the magnetic field component in the plane of the sky. The WISE image traces the emission of PAHs, the brightness of which scales linearly with the product of the intensity of the radiation field and the gas column density \citep{Draine07,Compiegne11}.
Measurements of these observational tracers computed at the center positions of the four JWST images and for the surrounding background are listed in Table~\ref{tab:Planck_dust}. The background measurements correspond to the median value calculated for pixels within $2^\circ$ of Merope where $\tau_{353}$ is smaller than $2\times 10^{-5}$ to exclude the Pleiades area. 

\begin{figure}[ht]
    \centering
    \includegraphics[width = 250 pt]{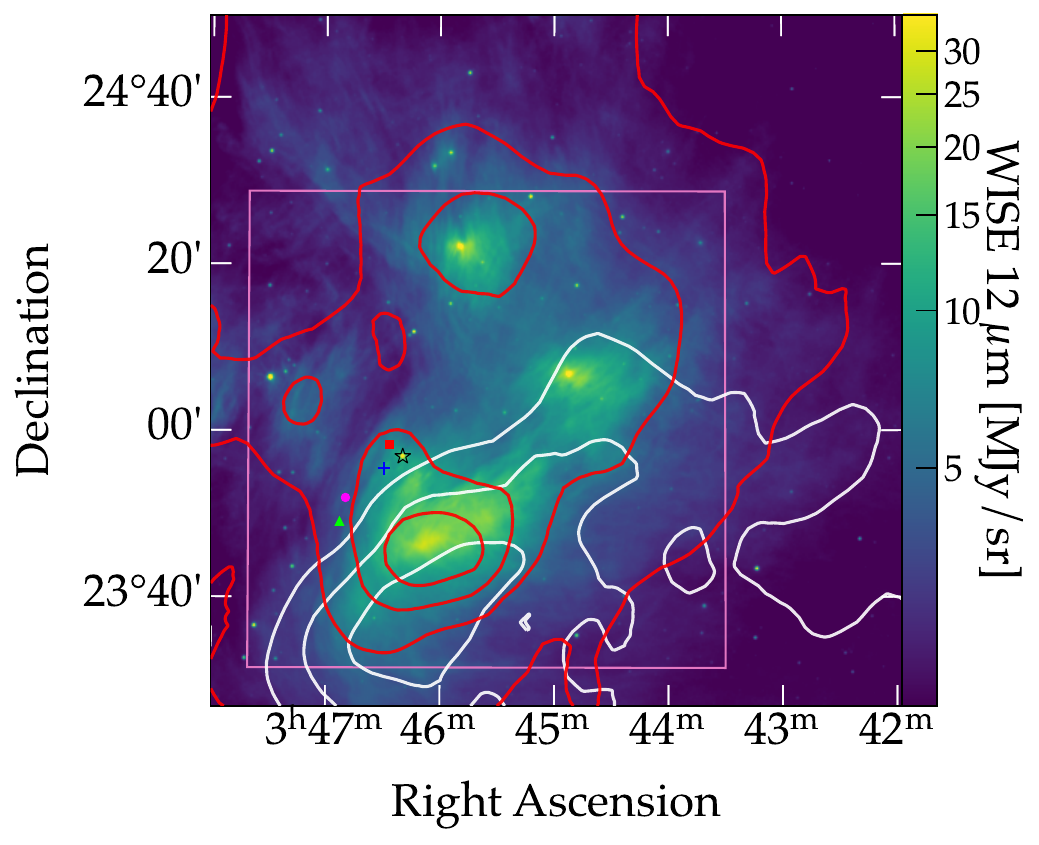}
    \caption{Dust temperature and opacity maps overlaid on  the $12\,\mu $m  WISE image at full angular resolution. The dust opacity at 353\,GHz and the dust temperature, as derived from the analysis of {\it Planck} data, are shown with white contours for $\tau_{353} = 2$, 3, and $4\times 10^{-5}$ in and red contours for $T_d = 20$, 23, 26, and  29\,K. Merope is marked with a star, the centers of the JWST images with a red square,  Field~1 Cuts~1 and 2 with a blue plus sign, and Field~2 Cuts~1 and 2 with a purple circle and a green triangle. The pink box displays the location of the image of dust polarization angles presented in Fig.~\ref{fig:Pleiades_Planck_polar}.}
    \label{fig:Pleiades_WISE_Planck}
\end{figure}

Figure~\ref{fig:Pleiades_WISE_Planck}  shows the \textit{Planck} maps of dust temperature $T_d$ and submillimeter opacity map $\tau_{353}$ at $353\,$GHz overlaid on the $12\,\mu$m WISE image at 12\arcsec resolution\footnote{The image was downloaded from \url{https://irsa.ipac.caltech.edu/applications/wise}}. At the distance of the Pleiades stars, the angular size of the image corresponds to 3\,pc. The dust temperature exceeds the background value across the full extent of the mid-infrared emission from the nebula.  The opacity contours delineate an interstellar cloud to the south of Merope. It corresponds to a small cloud identified by \citet{Ungerechts87} in their  survey with a peak CO(1-0) brightness of $6.6\,\mathrm{K\,km\,s^{-1}}$. 
The Merope star and the two JWST fields are located at the edge of this translucent cloud. The overlap between the contours of $T_d$ and $\tau_{353}$ indicates that the cloud is locally heated by Merope. 

\begin{figure}[ht]
    \centering
    \includegraphics[width = 250 pt]{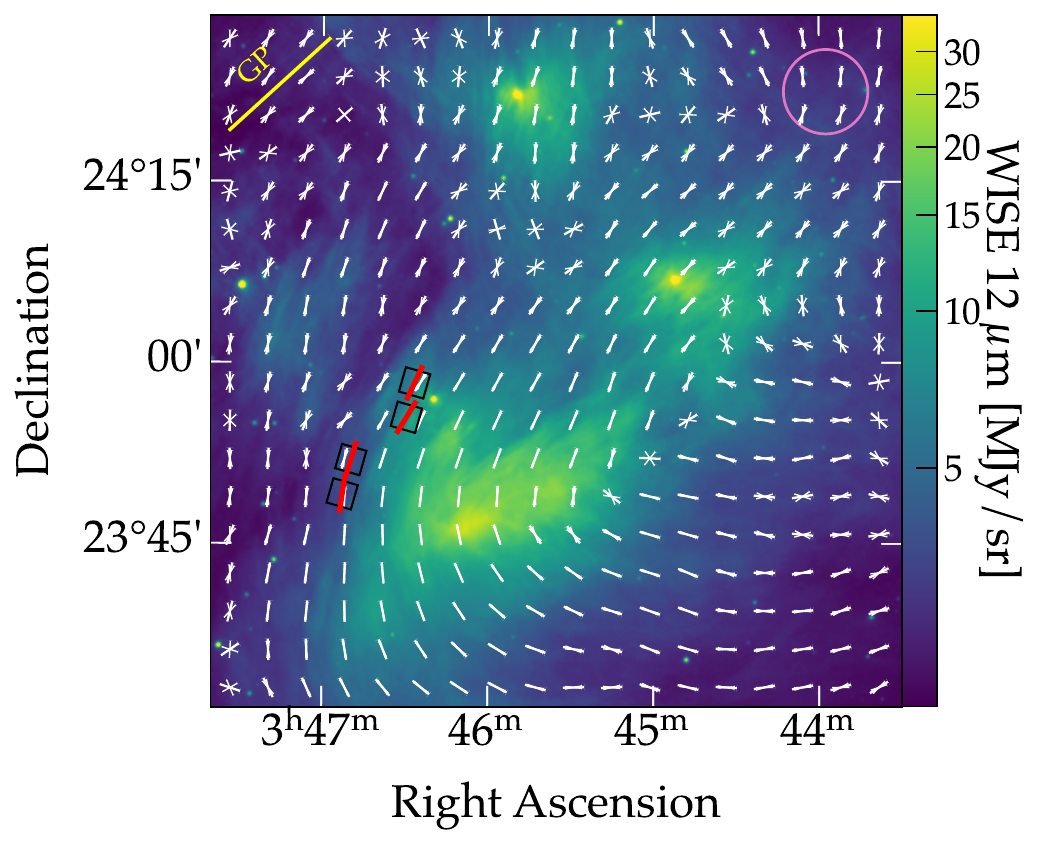}
    \caption{\textit{Planck} dust polarization angles rotated by $90^\circ$ to display the magnetic field orientation on the plane of the sky.  
    The white segments indicating the dust polarization angle are plotted alongside thinner adjacent segments that represent $1 \sigma$ error bars. The background image is the same $12\,\mu $m WISE image at full resolution as in Fig.~\ref{fig:Pleiades_WISE_Planck}. 
    The circle in the top-right corner shows the beam of the {\it Planck} data at half maximum. The black squares indicate the JWST image fields, with the red segments showing the orientations of the striations in the JWST maps. The yellow line in the top-left corner shows the orientation of the Galactic longitude axis.}
    \label{fig:Pleiades_Planck_polar}
\end{figure}

Figure~\ref{fig:Pleiades_Planck_polar} shows polarization angles, which have been  rotated by $90^\circ$ to display the magnetic field orientation in the plane of the sky. 
Angles were computed from the {\it Planck }  maps after the background contribution to the Stokes $Q$ and $U$ maps was subtracted. Therefore, the figure illustrates the orientation of the magnetic field within the Pleiades nebula. It shows how the magnetic field changes orientation over the nebula, in particular over the cloud to the south of Merope.
The black squares indicate the JWST image fields, and the red segments show the orientations of the striations (the angles $\theta_0$ listed in Table~\ref{tab:angles}). Figure~\ref{fig:Pleiades_Planck_polar} shows a good match between the red and white segments. The polarization angles $\psi$ measured at the center position of the JWST images can be found in Table~\ref{tab:angles}.

\section{Association with the cold neutral medium}
\label{sec:context}

In the paper introduction, we claim that the JWST Pleiades observations probe the structure of the CNM within a few parsecs of Merope. To support and discuss this statement, we used the {\it Planck} and WISE data presented in Appendix~\ref{App:interstellar}. These data provide a global perspective on the JWST observations. The $5^\prime$ angular resolution of the {\it Planck} data corresponds to a scale of 0.2\,pc at the distance of Merope. 

Merope lies to the front of the large (150 pc in diameter) Per-Tau bubble that was identified by \citet{Bialy21} using 3D stellar reddening data. The star is close to the outer surface of the shell that surrounds the bubble. On smaller scales, the dust emission data presented in Appendix~\ref{App:interstellar}  show that Merope and the two JWST fields are located at the edge of a translucent cloud that has a diameter of a few parsecs. Table~\ref{tab:Planck_dust} lists the values of the dust temperature $T_d$ and the dust opacity $\tau_{353}$ in the center positions of the four JWST images, as well as for the surrounding background. The values of the {\it Planck} Stokes parameters at 353\,GHz, and the PAH emission at $12\,\mu$m, measured on the WISE map of the entire sky produced by \citet{Meisner14}, are also listed in the table. We used these data to estimate the radiation field intensity and gas column density at the location of the JWST fields. 

The radiation field intensity can be estimated from the dust temperature using the formula
\begin{equation}
   G = (T_d/T_\mathrm{BG})^{(4+\beta)} ,
   \label{eq:Gfac}
\end{equation}
with a spectral index $\beta = 1.6$ \citep{Planck14_XI}. The factor $G$ measures the radiation field intensity normalized to its value within the background ISM, away from the Pleiades stars.  This formula yields $G$ values ranging from 6 to 9 at the position of the JWST fields. 
The opacity values listed in Table~\ref{tab:Planck_dust} include the Galactic background. To  estimate the column density of the cloud associated with Merope,  we needed to subtract this contribution. This subtraction introduces a large uncertainty equal to the dispersion of opacity values measured over the background area, which is listed as the error bar in the background line of the table. Taking into account this uncertainty, we assumed that the column density of the cloud at the positions of the JWST fields cannot be much larger than $5\times 10^{20}\,\mathrm{H\,cm}^{-2}$. 

The $12\,\mu$m brightness allows us to crosscheck the estimates of $G$ and  $N_H$. In the solar neighborhood, the mean emission at $12\,\mu$m is $0.3\,\mathrm{MJy\,sr^{-1}}$ for a column density $N_H$ of $10^{21}\,\mathrm{H\,cm}^{-2}$ \citep{Compiegne11}. 
For fixed PAH abundance and emission properties, this emission scales linearly with the product $G \times N_H$ \citep{Draine07,Compiegne11}.  Based on the $12\,\mu$m brightness values in Table~\ref{tab:Planck_dust}, we estimated the product $G \times N_H$ to be 16 and 7 for Fields~1 and 2, where $N_H$ is expressed in units of $10^{21}\,\mathrm{H\,cm}^{-2}$. These values suggest that the estimates of G and $N_H$ derived from dust temperature and opacity are somewhat underestimated.
This could be due to beam smoothing and the superposition of cloud and background emissions, which were not considered when fitting the dust temperature and opacity. 
In any case, the
estimates of $G$ and $N_H$ support our assertion that the JWST observations characterize emission from CNM within a few parsecs of Merope. The value of $G$ is 10 at a distance of 0.5 \,pc from Merope. 
The contribution to $N_H$  of warm gas with densities of a few tenths of  a $\mathrm{H\,cm^{-3}}$ across this distance is negligible.

We also estimated the contribution of the Galactic background to the JWST power spectra. 
To achieve this, we used the scaling law established by \citet{Miville07} when analyzing Infrared Astronomical satellite (IRAS) data. The fraction of power contributed by the Galactic background $f^\mathrm{BG}$  is estimated using the following formula:
\begin{equation}
    f^\mathrm{BG} \, = (I_{3.5}^\mathrm{BG}/I_{3.5})^2,
\end{equation}
where $I_{3.5}$, the brightness in the center of a JWST image in the F335M filter, is obtained by multiplying the values $I_{12}$ and $I_{3.5}/I_{12}$ in Table~\ref{tab:coordinates}.
To obtain the  background brightness $I_{3.5}^\mathrm{BG}$ in the F335M filter, we scaled the background brightness at $12\,\mu$m listed in Table~\ref{tab:Planck_dust} by the intensity ratio $6\,10^{-2}$, as determined in the study of \citet{Flagey06}. 
We obtain values of $f^\mathrm{BG}$ $\sim 0.1$\% for the Field~1 images and between one and a few percent for the Field~2 images. These low values support the link that the paper makes between the JWST power spectra and the CNM near the Pleiades stars.

\end{document}